\documentclass[aps,prd,reprint,groupedaddress,onecolumn,showpacs]{revtex4-1}
\usepackage{color}
\usepackage{graphicx}% Include figure files
\usepackage{bm}% bold math
\usepackage{amssymb,amsmath,amsfonts, mathrsfs}

\usepackage{url}
\begin{document}
\begin{flushright}USTC-ICTS-16-15
\end{flushright}
\title {On Friedrichs Model with Two  Continuum States}
\author{Zhiguang Xiao}
\email[]{xiaozg@ustc.edu.cn}
%\thanks{}
\affiliation{Interdisciplinary Center for Theoretical Study, University of Science
and Technology of China, Hefei, Anhui 230026, China}

\author{Zhi-Yong Zhou}
\email[]{zhouzhy@seu.edu.cn}
%\thanks{}
\affiliation{Department of Physics, Southeast University, Nanjing 211189,
P.~R.~China}
\affiliation{
Kavli Institute for Theoretical Physics China, CAS, Beijing 100190, China}


\date{\today}

\begin{abstract}
The Friedrichs model with one discrete state coupled to more than one
continuum is studied. The exact eigenstates for the full Hamiltonian
can be solved explicitly. The discrete state is found to generate more
than one virtual state pole or more than one pair of resonance poles
in different Riemann sheets in different situations.  The form factors
could also generate new states on different sheets. All these states
can appear in the generalized completeness relation.
\end{abstract}

%\pacs{12.39.Jh, 13.20.Fc, 13.75.Lb, 11.55.Fv}
%\keywords{}

\maketitle
\section{Introduction}
Due to the intense experimental activities, more and more heavy
quarkonium-like states were found these years in hadron spectroscopy
as listed in the Particle Data Group Table~\cite{Agashe:2014kda}.
However, many hadron states near or above the open-flavor thresholds
can hardly be explained in the conventional ``quenched"  potential
models, in which a meson state is regarded as the bound state of a
quark and an anti-quark with a coulomb potential term at short
distance and a linear confinement potential term at large distance,
such as the Godfrey-Isgur (GI) model~\cite{Godfrey:1985xj}.
The discrepancies between the experiments and the potential model
calculations seem to be made up by considering the couplings of the
quark-antiquark bound states and the continuum hadronic states in
different realizations, such as the
coupled-channel model~\cite{Pennington:2007xr,Coito:2012vf,Zhou:2013ada}, and the
screened potential model~\cite{Li:2009zu}.
This mechanism also plays an
important role when a discrete state is coupled strongly with more than
one continuum states. Typically, the enigmatic $\sigma$ and $\kappa$
resonances in $\pi\pi$ and $\pi K$
scatterings~\cite{Xiao:2000kx,Zheng:2003rw} can hardly be accommodated
in the conventional quark model, because they are strongly coupled
with the continuum states. In Ref.~\cite{zhou:2010ra}, most  of the
$0^+$ meson states below 2 GeV are found to be related to the
resonance poles generated by a few bare states coupled with several
thresholds, and some of the resonant states share the same origin of
the bare state. In Ref.~\cite{zhou:2011sp,Zhou:2013ada}, some of the
charmed, charmed-strange, and charmonium-like states are discussed in
the same spirit, where the inclusion of the hadron-loop effects causes
the masses and widths of the states near or above the thresholds
shifting from the GI's prediction closer to the experimental data.
With these coupling of the bare states and the continuum states, one
would expect that the discrete bare states are no longer the
eigenstate of the Hamiltonian and the resonance found in the
experiments should be a linear superposition of the original discrete
state and the continuum states.  It would be desirable if one could
describe the wave function of the resonance in terms of the discrete
state and the continuum states.

 The Friedrichs model~\cite{Friedrichs:1948,horwitz1971} is a kind of solvable model which couples a
discrete state  to a continuum state.
The generalized eigenstates of the full interacting
Hamiltonian, can be solved explicitly in terms of the original
discrete state and the continuum states.  The discrete
state may become unstable and be described by Gamow states. These
states can not be described by the vectors of the usual Hilbert space but by the
ones in Rigged Hilbert Space (RHS)\cite{Antoniou1993443}. RHS is composed of a Gel'fand
triplet $\Omega\subset \mathscr H \subset \Omega^\times$, where
$\mathscr H$ is the usual Hilbert space of the normalizable states,
$\Omega$ is a nuclear space which is dense in $\mathscr H$, and
$\Omega^\times$ is the space of the anti-linear continuous functionals
on the nuclear space. Gamow states  must be in the larger $\Omega^\times$,
since it is the generalized eigenstate of the full Hamiltonian with
complex eigenvalues. The descriptions of the in-states and the
out-states are
using different Rigged Hilbert spaces, $\Omega_{\pm}\subset \mathscr H
\subset \Omega_\pm^\times$, where the subscript ``$-$" denotes the out-state
space and ``$+$" denotes the in-state space. The triplet of state spaces
can be mapped to the complex function spaces $D_{\mp}\subset \mathscr
H^2_\mp\subset D_{\mp}^\times $ where $D_{\mp}=S\cap \mathscr
H^2_{\mp}|_{\mathbb R^+}$, respectively, to form a representation. $S$
is the Schwarz space and  $\mathscr H^2_{\mp}$ is the so-called Hardy
space in which the functions are analytic on  $\mathbb C_\mp$ and
$|_{\mathbb R^+}$ means restriction on $\mathbb R^+$. There are also
two kinds of Gamow states, $|z_R^-\rangle\in \Omega_-^\times$,
$|z_R^+\rangle\in \Omega_+^\times$ denoting the decaying state and
growing states which correspond to the lower and upper second sheet
poles of the $S$-matrix, respectively. For further detailed
discussions
on the mathematical foundation, the readers are referred to~\cite{Bohm:1989,Gadella:2004}.

In the original Friedrichs model where only one discrete and one
continuum state are involved, the number of the discrete state poles
generated from the original discrete state is doubled on the
two-sheeted
Riemann surface of the analytically continued $S$-matrix. If the
energy  of the original discrete state is below the threshold, it will generate
a bound state pole on the first sheet and a virtual state pole on the
second sheet of the Riemann surface~\cite{Xiao:2016dsx}. For the
original discrete state with its mass higher than the threshold,
it becomes a pair of resonance poles on the second sheet for small
coupling.  There can also be poles
generated by the  the form factor~\cite{Likhoded:1997}. In our previous
paper~\cite{Xiao:2016dsx}, we gave a general argument that for each simple pole of the form
factor, there will be a second-sheet pole of the $S$-matrix generated
from this pole.
The similar argument can also be  applied to the exponential form
factor and it can be demonstrated that there should be a virtual state pole generated
from the minus infinity.

Some extensions of the original Friedrichs model to more than one
system are made, most of which deals with more discrete states
coupled with a continuum such as in Refs.
\cite{STEY19721,Bailey1978,PhysRevA.70.032702,Antoniou2003,Gadella:2011mj}.
It is also extended to include the fermion-boson
interactions~\cite{Civitarese:2007zj,Civitarese:2007zk}.
In the present paper, we will extend
the Friedrichs model to describe the process of coupling one discrete
state with more than one  continuum states with different thresholds, which is more relevant to
the phenomenological studies of hadron spectroscopy, such as the
hadron-loop models and the coupled-channel models.
We mainly consider the scenario of
including two continua, and then it is straightforward to generalize
the result to the cases with more than two
continua. The explicit solutions of the Friedrichs model with
one discrete state coupled to two continua are presented, which is not found in the literature. Furthermore,
in the two-continuum Friedrichs model, since the analytically continued
$S$-matrix is defined on  a four-sheeted Riemann surface, the number of the  discrete
state poles will be doubled twice. For small couplings, we will
discuss on which sheet the poles could be located in different
conditions. These poles could be bound states on the first sheet or
virtual states and resonances on the other sheets. All these states can be expressed as the superposition of the
original discrete state and the continuum states. From the
mathematical point of view, as in the single-continuum case,
the singularity of the form factor can also generate
states on different sheets.

In the original formulation of Friedrichs,
only the continuum and the bound states could enter the completeness
relation. However, Petrosky, Prigogine and Tasaki
(PPT)~\cite{Prigogine:1991} proposed a way to define the continuum in terms of a kind of complex distribution, in
which the prescripted contour information of the energy are encoded in the continuum
right eigenstates. In this formulation, the states corresponding to the
unphysical sheet poles could also enter the completeness relation. We
also generalize this kind of formulation to the multi-continuum cases,
in which all the bound states, virtual states,
and resonances are included in the completeness relation equally.

We organize the paper as follows: Section \ref{sect:solu} gives the
continuum solution to the Friedrichs model with one discrete state and two
continua. Section \ref{sect:polepos} discusses the possible pole
positions for small couplings. Section \ref{sect:Gamow} discusses the
discrete states solutions in this model and the completeness relation.
Section \ref{sect:general} generalizes the results to Friedrichs
models with more than two continua. Section \ref{sect:conclude} devotes
to the conclusion and discussions.

\section{Solution to Friedrichs model with one discrete state and two
continua \label{sect:solu}}

Suppose the full Hamiltonian is $H=H_0+V$ in which the free
Hamiltonian $H_0$ has one discrete eigenstate $|1\rangle$ and two
kinds of continuum eigenstates, $|\omega\rangle_1$
and $|\omega\rangle_2$, that is,
\begin{eqnarray}
H_0|1\rangle=\omega_0|1\rangle,\nonumber\\
H_0|\omega\rangle_{1}=\omega|\omega\rangle_{1},\nonumber\\
H_0|\omega\rangle_{2}=\omega|\omega\rangle_{2}.
\end{eqnarray}
 The continuum states are coupled to the discrete
state $|1\rangle$ with  the coupling strength denoted by $\lambda_{1}$
and $\lambda_{2}$ respectively, and there
is no direct interaction between the two
continua. The interaction term of the Hamiltonian can be expressed as
\begin{eqnarray}
V&=&\lambda_1\int_{a_1}^\infty\mathrm{d}\omega[f_1(\omega)|\omega\rangle_1\langle
1|+f_1^*(\omega)|1\rangle{\ }_1\langle\omega|]\nonumber\\
&+&\lambda_2\int_{a_2}^\infty\mathrm{d}\omega[f_2(\omega)|\omega\rangle_2\langle
1|+f_2^*(\omega)|1\rangle{\ }_2\langle\omega|] \,,
\end{eqnarray}
where $a_1$ and $a_2$ are the thresholds for the two continua  in the energy representation with
$a_2>a_1$. Suppose the eigenstate $|\Psi(x)\rangle$ of $H$ is expressed as
\begin{eqnarray}
|\Psi(x)\rangle=\alpha(x)|1\rangle+\int_{a_1}^\infty
\psi(x,\omega)|\omega\rangle_1\mathrm{d}\omega+\int_{a_2}^\infty
\phi(x,\omega)|\omega\rangle_2\mathrm{d}\omega.  \label{eq:ansatz-Phi}
\end{eqnarray}
Then, a group of equations are obtained from the eigenequation
$H|\Psi(x)\rangle=x|\Psi(x)\rangle$,
\begin{align}
(\omega_0-x)\alpha(x)+\lambda_1\int_{a_1}^\infty
f_1^*(\omega)\psi(x,\omega)\mathrm{d}\omega+\lambda_2\int_{a_2}^\infty
f_2^*(\omega)\phi(x,\omega)\mathrm{d}\omega=0,&&\label{eq:alpha}\\
(\omega-x)\psi(x,\omega)+\lambda_1\alpha(x)f_1(\omega)=0,&&(\text{ for
}\omega>a_1),\label{eq:psi}\\
(\omega-x)\phi(x,\omega)+\lambda_2\alpha(x)f_2(\omega)=0,&&(\text{ for
}\omega>a_2).\label{eq:phi}
\end{align}
To solve these equations, we distinguish three cases.

\noindent {\it Case a}) When  $x\notin [a_1,\infty]$, the solutions to
Eqs. (\ref{eq:psi})
and (\ref{eq:phi}) are
\begin{eqnarray} \psi(x,\omega)&=&
\frac{\lambda_1\alpha(x)f_1(\omega)}{x-\omega},\label{eq:ansatz-psi-0}\\
\phi(x,\omega)&=& \frac{\lambda_2\alpha(x)f_2(\omega)}{x-\omega}.
\label{eq:ansatz-phi-0}
\end{eqnarray}
Inserting them into
Eq.(\ref{eq:alpha}), we have the equation
\begin{align}
\eta(x)=x-\omega_0-\lambda_1^2\int_{a_1}^\infty\frac{G_1(\omega)}{x-\omega}\mathrm{d}\omega-\lambda_2^2\int_{a_2}^\infty\frac{G_2(\omega)}{x-\omega}\mathrm{d}\omega =0
\label{eq:eta-c}
\end{align}
where $G_1(\omega)=f_1(\omega)f_1^*(\omega)$ and
$G_2(\omega)=f_2(\omega)f_2^*(\omega)$. If this equation has solution
$x<a_1$ on the real axis, the corresponding
eigenvector will represent a bound state which is the renormalized state
of the discrete unperturbed state $|1\rangle$.
If this
happens, since the two integrals are negative, $x$ must be less than
$\omega_0$, which means that the interaction pulls down the
energy of  the discrete state.
A necessary condition for this to happen is
\begin{align}
\omega_0<a_1+\lambda_1^2\int_{a_1}^\infty\frac{f_1(\omega)f_1^*(\omega)}{\omega-a_1}\mathrm{d}\omega+\lambda_2^2\int_{a_2}^\infty\frac{f_2(\omega)f_2^*(\omega)}{\omega-a_1}\mathrm{d}\omega\,.
\end{align}
If $\omega_0<a_1$ it will guarantee that the system has a bound state.
Moreover, the $\eta(x)$ function can be analytically continued to a
four-sheeted Riemann surface since it has two cuts. We will see that
$\eta(x)$ may also have zeroes on the second, third, and fourth sheets. In
these cases, these zero points are complex generalized eigenvalues on different Riemann
sheets and the corresponding generalized eigenstates can also be obtained.
From Eq.(\ref{eq:ansatz-Phi}), the
eigenstate is recast into
\begin{eqnarray}
|\Psi_0(x)\rangle=\alpha(x)\Big[|1\rangle+\lambda_1\int_{a_1}^\infty \mathrm
d\omega
\frac{f_1(\omega)}{x-\omega}|\omega\rangle_1+\lambda_2\int_{a_2}^\infty
\mathrm d\omega \frac{f_2(\omega)}{x-\omega}|\omega\rangle_2\Big]\,.
\label{eq:solu-Psi-0} \end{eqnarray}
These are generalized eigenstates with discrete eigenvalues which are
zero points of the analytically continued $\eta(z)$.
For bound states, the normalization can be chosen as
$\alpha(x)=(1/\eta'(x))^{1/2}$ such that
$\langle\Psi_0(x)|\Psi_0(x)\rangle=1$.
We will discuss more about these discrete states later.

\noindent {\it Case b}) When
$a_1<x<a_2$, since Eq.(\ref{eq:phi}) only exists in $\omega>a_2$, only
$\psi$ can have $\delta$ function contribution
\begin{eqnarray}
\psi_\pm(x,\omega)&=&
\frac{\lambda_1\alpha_\pm(x)f_1(\omega)}{x-\omega\pm
i0}+\gamma_{1\pm}(\omega)\delta(\omega-x),\label{eq:ansatz-psi-1}\\
\phi_\pm(x,\omega)&=&
\frac{\lambda_2\alpha_\pm(x)f_2(\omega)}{x-\omega}\,.
\label{eq:ansatz-phi-1}
\end{eqnarray}
We have inserted the $i 0$ in the denominator to avoid the
singularity at $x=\omega$. Inserting them into
(\ref{eq:alpha}), we then have
\begin{align}
(\omega_0-x)\alpha_\pm(x)+\lambda_1
f_1^*(x)\gamma_{1\pm}(x)+\alpha_\pm (x)\lambda_1^2\int_{a_1}^\infty
\frac{|f_1(\omega)|^2}{x-\omega\pm i0
}\mathrm{d}\omega+\alpha_\pm{ (x)}\lambda^2_2\int_{a_2}^\infty \frac
{|f_2(\omega)|^2}{x-\omega}\mathrm{d}\omega=0\,. \label{eq:alpha-1}
\end{align}
If we define
 \begin{eqnarray}
\eta^\pm(x)=x-\omega_0-\lambda_1^2\int_{a_1}^\infty\frac{G_1(\omega)}{x-\omega\pm
i
0}\mathrm{d}\omega-\lambda_2^2\int_{a_2}^\infty\frac{G_2(\omega)}{x-\omega\pm
i 0}\mathrm{d}\omega\,,
\label{eq:eta-coupled}
\end{eqnarray}
$\gamma_{1\pm}$  can be expressed as
\begin{align}\gamma_{1\pm}(x)=\frac
{\alpha_\pm(x)\eta^{\pm}(x)}{\lambda_{1} f_1^*(x)}\,.
\end{align}
With the choice of the normalization $
\alpha_\pm=\frac{\lambda_{1}f^*_1(x)}{\eta^{\pm}(x)}$ such that  $\langle
 \Psi_{1\pm}(x)|\Psi_{1\pm}(x')\rangle=\delta(x-x')$, the eigenstate
for eigenvalue $x$ can then be obtained
\begin{align}
|\Psi_{1\pm}(x)\rangle=|x\rangle_1+\frac{\lambda_1
f_1^*(x)}{\eta^\pm(x)}\Big[|1\rangle+\lambda_1\int_{a_1}^\infty \mathrm
d\omega \frac{f_1(\omega)}{x-\omega\pm
i0 }|\omega\rangle_1+\lambda_2\int_{a_2}^\infty \mathrm d\omega
\frac{f_2(\omega)}{x-\omega\pm i 0}|\omega\rangle_2\Big]\,.
\label{eq:solu-final-1}
\end{align}
In Eqs. (\ref{eq:eta-coupled}) and  (\ref{eq:solu-final-1}) the $\pm
i 0$'s in
the last integrals has no effect for $x<a_2$, but we keep them in order
to extend these equations to $x>a_2$ case below.

\noindent {\it Case c}) Let us now look at $x>a_2$. There are two
degenerate states for the free Hamiltonian. We would expect that there
are also two degenerate eigenstates of the full Hamiltonian for these eigenvalues. From
 Eq.~(\ref{eq:psi}) and  Eq.~(\ref{eq:phi}),  Eq.~(\ref{eq:ansatz-psi-1}) is not changed
and Eq. (\ref{eq:ansatz-phi-1}) becomes
\begin{align} \phi_\pm(x,\omega)=
\gamma_2(\omega)\delta(\omega-x)-\frac{\lambda_2\alpha_\pm(x)f_2(\omega)}{\omega-x\pm
i0}\,.
\label{eq:ansatz-phi-2}
\end{align}
It is obvious that the solution of {\it Case b}) is also the solution
for this case with
$\gamma_2=0$.   If the couplings are turned off, this solution goes back to
$|x\rangle_1$.
 There must be another solution which becomes
$|x\rangle_2$ as $\lambda_{1,2}\to 0$. We start with ansatz
(\ref{eq:ansatz-phi-2}) and set $\gamma_1=0$ in
(\ref{eq:ansatz-psi-1}), and then have
\begin{align}
\psi_\pm(x,\omega)=& -\frac{\lambda_1\alpha_\pm(x)f_1(\omega)}{\omega-x\pm
i0}\,.
\end{align}
Similar to Eq. (\ref{eq:solu-final-1}), after choosing the suitable
normalization $\alpha_\pm(x)=\frac{\lambda_2 f_2^*(x)}{\eta^\pm(x)}$ such that
$\langle\Psi_{2\pm}(x)|\Psi_{2\pm}(x')\rangle=\delta(x-x')$, we then obtain the other
solution for eigenvalue $x>a_2$,
\begin{align}
|\Psi_{2\pm}(x)\rangle=|x\rangle_2+\frac{\lambda_2
f_2^*(x)}{\eta^\pm(x)}\Big[|1\rangle+\lambda_1\int_{a_1}^\infty \mathrm
d\omega \frac{f_1(\omega)}{x-\omega\pm
i0}|\omega\rangle_1+\lambda_2\int_{a_2}^\infty \mathrm d\omega
\frac{f_2(\omega)}{x-\omega\pm i0}|\omega\rangle_2\Big]\,.
\label{eq:solu-final-2}
\end{align}
It can be proved that $\langle \Psi_{1\pm} | \Psi_{2\pm}\rangle=0$.

In general, for eigenvalue $x>a_2$, the
eigenstate solution should be a superposition of
Eqs. (\ref{eq:solu-final-1}) and (\ref{eq:solu-final-2}).
$|\Psi_+(x)\rangle$ is
the in-state and $|\Psi_-(x)\rangle$ is  the out-state. The $\eta(x)$ function
defined in (\ref{eq:eta-c}) and $\eta^\pm$ in (\ref{eq:eta-coupled}) can be
analytically continued to one function defined on a four-sheeted Riemann surface, which we also denote as
$\eta(x)$, with $\eta^+$ and $\eta^-$
being the boundary functions on the upper rim and lower rim of the
cut on the first sheet. From {\it case a}), the poles of $1/\eta(x)$ on the different
Riemann sheets are the discrete eigenvalues and the corresponding
states may represent the bound states, virtual states, and
resonant states for the full Hamiltonian. Before studying these
solutions, we will first study the pole
behaviors for $1/\eta(x)$ for small couplings in the next section.

\section{Analysis of the pole positions for weak couplings
\label{sect:polepos}}
\begin{figure}
\begin{center}
\includegraphics[width=8cm]{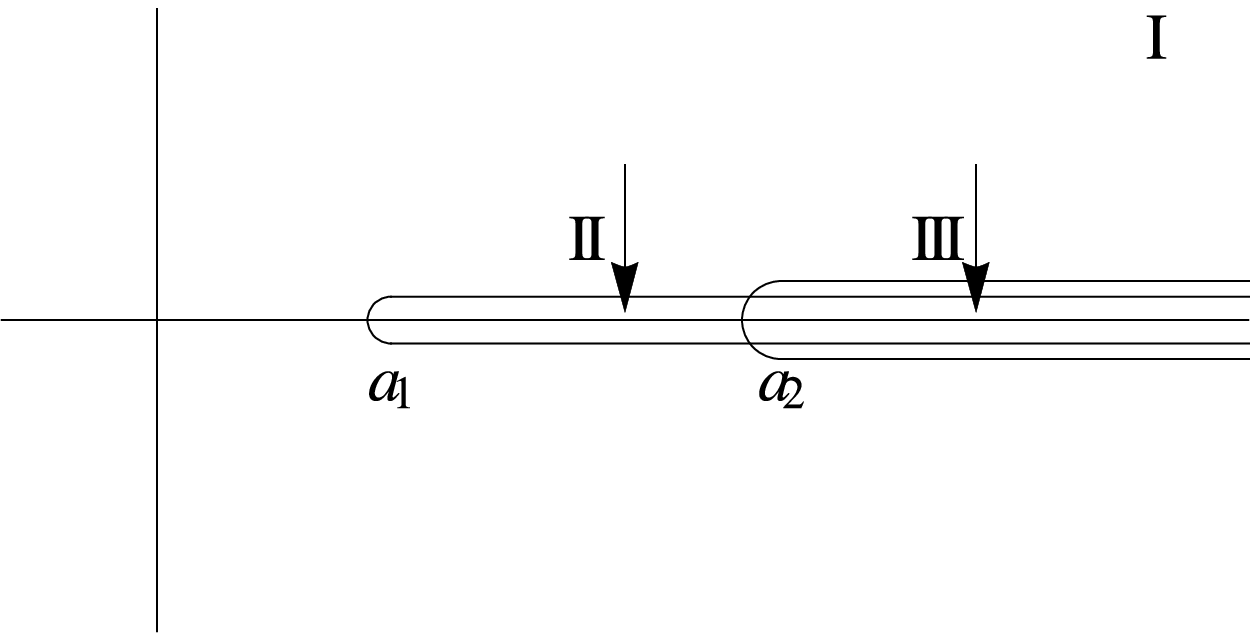}
\hskip 0.5cm
\includegraphics[width=8cm]{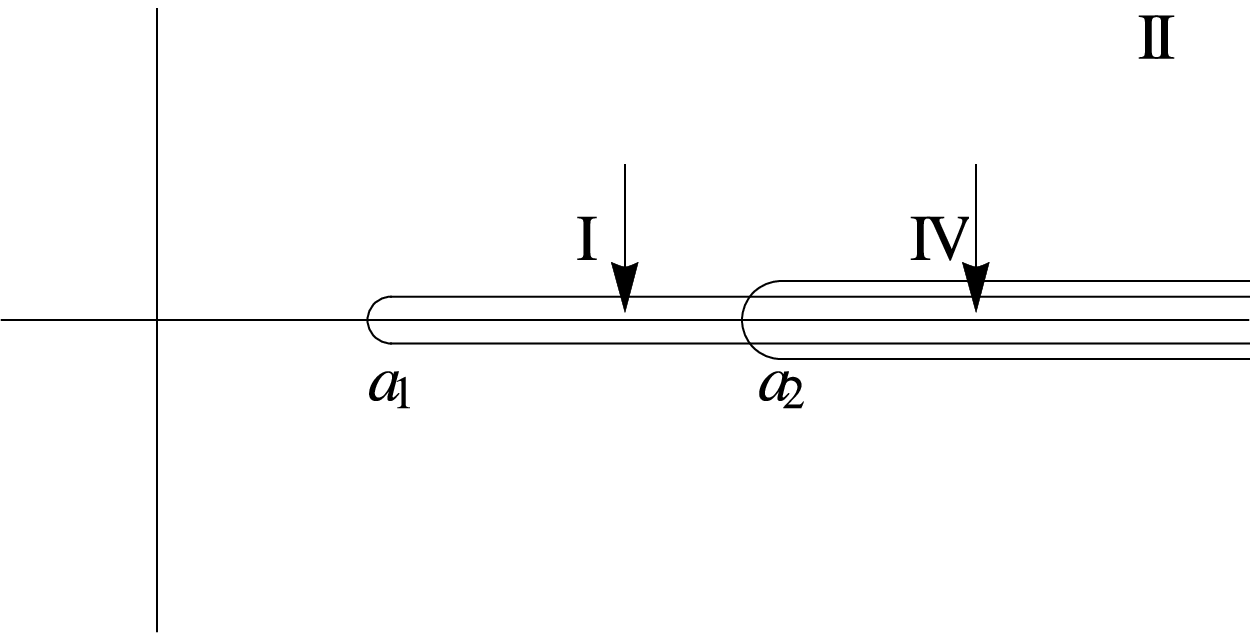}
\includegraphics[width=8cm]{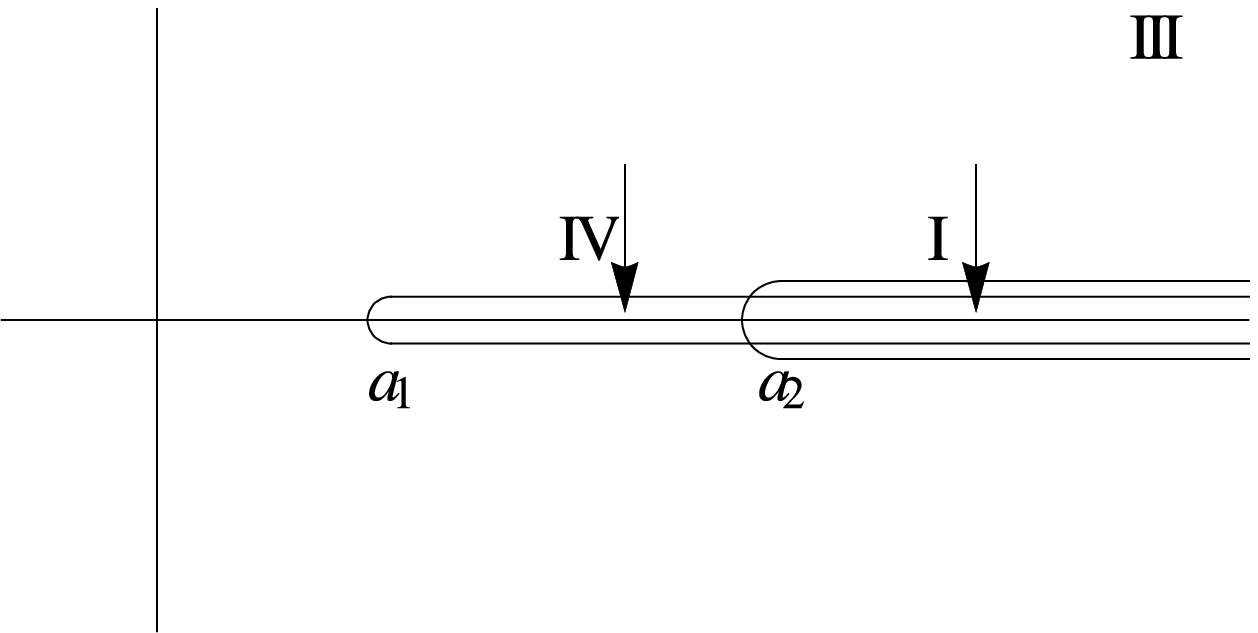}
\hskip 0.5cm
\includegraphics[width=8cm]{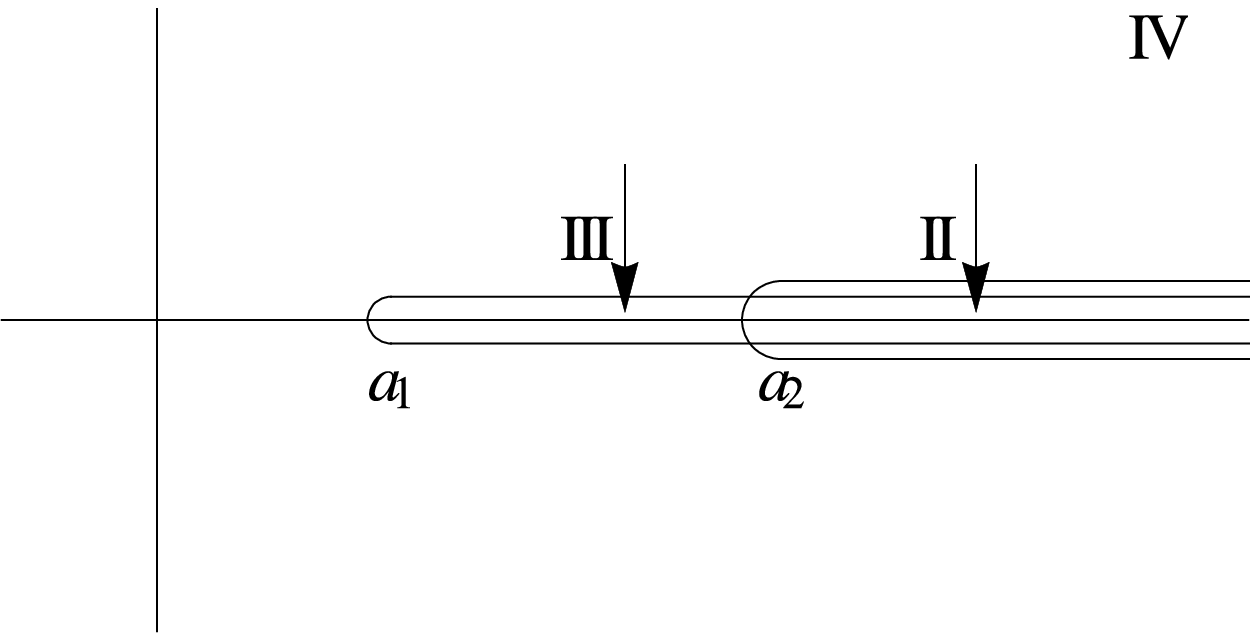}
\end{center}
\caption{Definition of the different Riemann sheets. $I$, $II$, $III$
and $IV$ on the up-right corner denote the current Riemann sheet for
each diagram. The Roman numerals beside the arrows mean the adjacent
Riemann sheets analytically continued from the current sheets through
the cut in the
direction of the arrows. \label {fig:RiemannSheet}}
\end{figure}

As in the  discussion of the simplest Friedrichs
model~\cite{Xiao:2016dsx}, we choose form factors
$G_1(\omega)=\frac {\sqrt{\omega-a_1}}{\omega +\zeta_1}$ and
$G_2(\omega)=\frac {\sqrt{\omega-a_2}}{\omega+\zeta_2}$ with
$\zeta_{1,2}>0,\zeta_{1,2}\in \mathbb R$, to illustrate the general behaviors
of the poles for small couplings. The two thresholds
for $\omega$ are at $a_1\ge0$ and $a_2>0$ with $a_2>a_1$.
 The analytically continued
$S$-matrix or $\eta(x)$ function will have four Riemann sheets and we define the second
sheet to be the one analytically continued from the cut between $a_1$ and
$a_2$ on the first sheet, the third sheet to be continued from the cut
above $a_2$ on the first sheet, and the fourth sheet to
be continued from the cut above $a_2$ on the second
sheet or from the cut between $a_1$ and $a_2$ on the
third sheet. See Fig.~\ref{fig:RiemannSheet} for  illustrations.

In the simplest Friedrichs model with one
continuum, it was shown that the
poles of the form factor always introduce poles of
$1/\eta(x)$ near these singularities on the unphysical sheet~\cite{Xiao:2016dsx}. Now there
are two form factors, we will see that the singularities of them will
also generate different poles on different
Riemann sheets.  From the definition of $\eta^\pm$,
Eq. (\ref{eq:eta-coupled}),
there are two integrals in the definition.  In each integral, the
form factor provides a pole to the integrand, which introduces  a pinch
singularity of $\eta$  at $-\zeta_1$ on both the second sheet and the third sheet for
the first integral, and a pinch singularity at $-\zeta_2$ on both the third
sheet and fourth sheet for the second integral. In fact, the $\eta(x)$
function on each sheet can be represented as,
\begin{align}
\eta^{II}(\omega)=&\eta^I(\omega)-2 \lambda_1^2 \pi i\, G_1(\omega),
\\
\eta^{III}(\omega)=&\eta^I(\omega)-2 \lambda_1^2 \pi i\, G_1(\omega)-2
\lambda_2^2 \pi i\, G_2(\omega),
\\
\eta^{IV}(\omega)=&\eta^I(\omega)-2 \lambda_2^2 \pi i\, G_2(\omega),
\end{align}
in which $G_1$ and $G_2$ are analytically  continued from the real
axis to the first Riemann sheet, and $I$, $II$, $III$, $IV$ label the
first, second, third, and fourth sheets, respectively. The analytically continued
$\eta(x)$ should satisfy the Schwartz reflection relation
$\eta^*(z)=\eta(z^*)$, which is the so-called real-analytic function. Thus,
the analytically continued
$G_{1,2}(z)$, which are proportional to the imaginary parts of the
integral,
should satisfy $G_{1,2}^*(z)=-G_{1,2}(z^*)$ on each sheet, which will
be called anti-real-analytic
function here. From above equations, the poles in $G_1$ and/or  $G_2$ in
each equation are just
the pinch singularities for the analytically continued $\eta (z)$ on
the corresponding sheet.

\begin{figure}
\includegraphics[height=2.2cm]{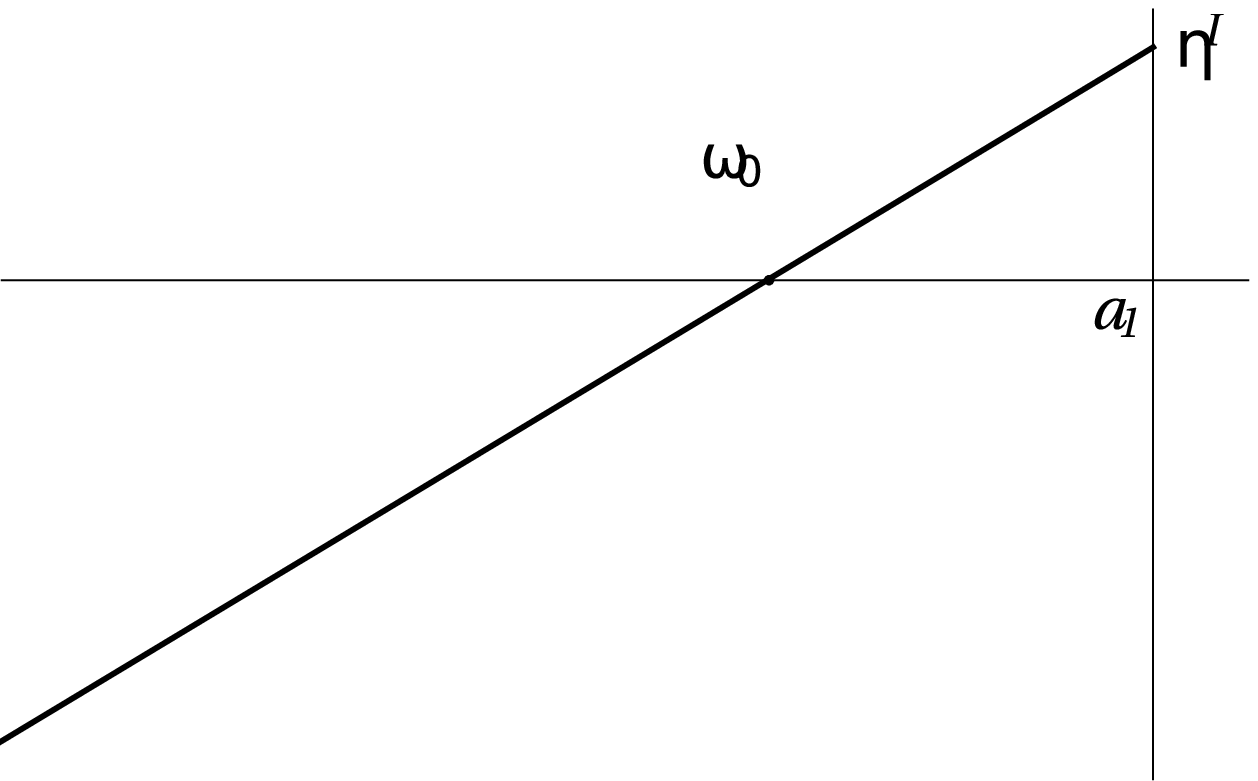}
\hspace{.5cm}
\includegraphics[height=2.2cm]{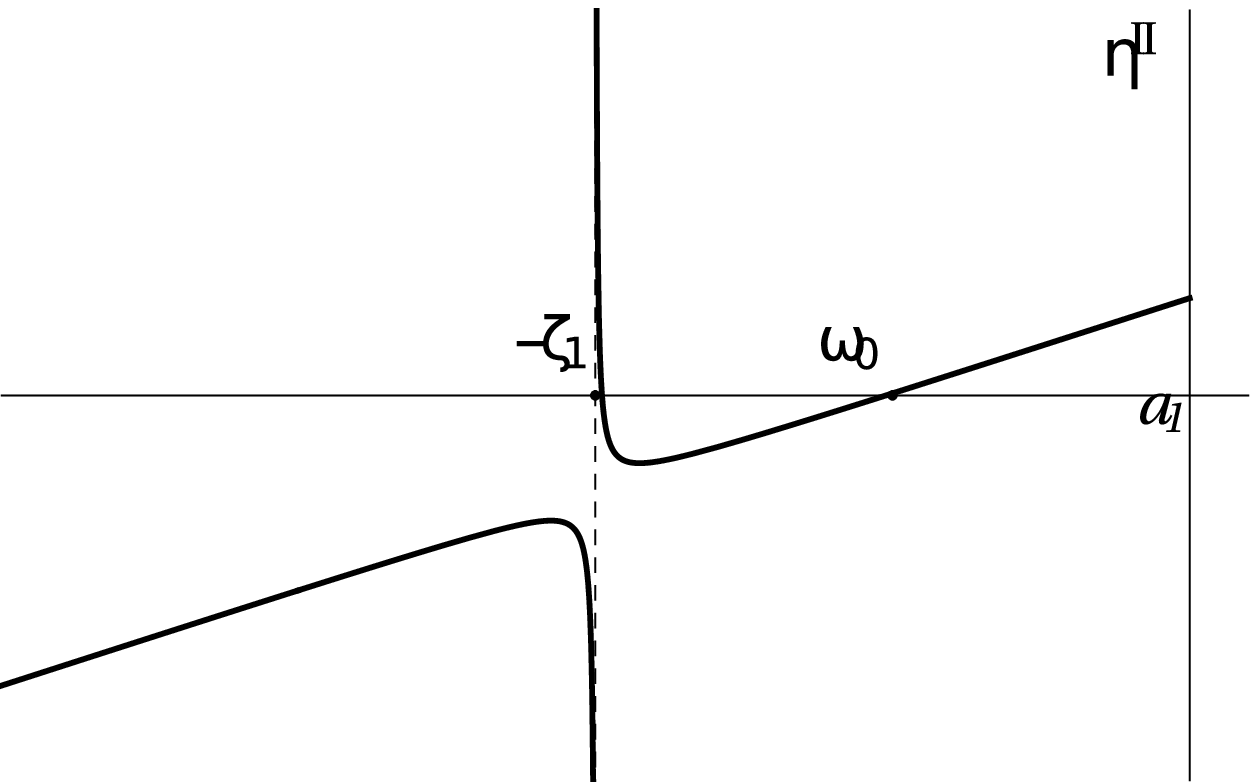}
\hspace{.5cm}
\includegraphics[height=2.2cm]{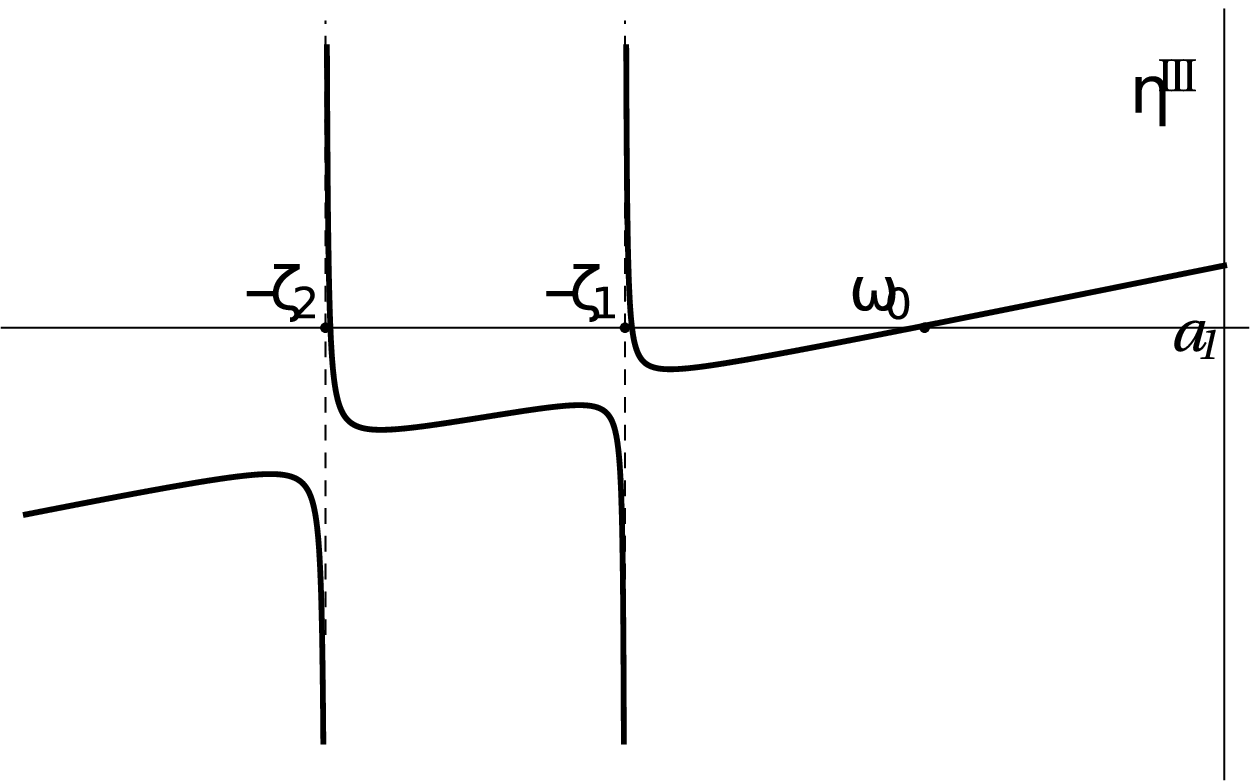}
\hspace{.5cm}
\includegraphics[height=2.2cm]{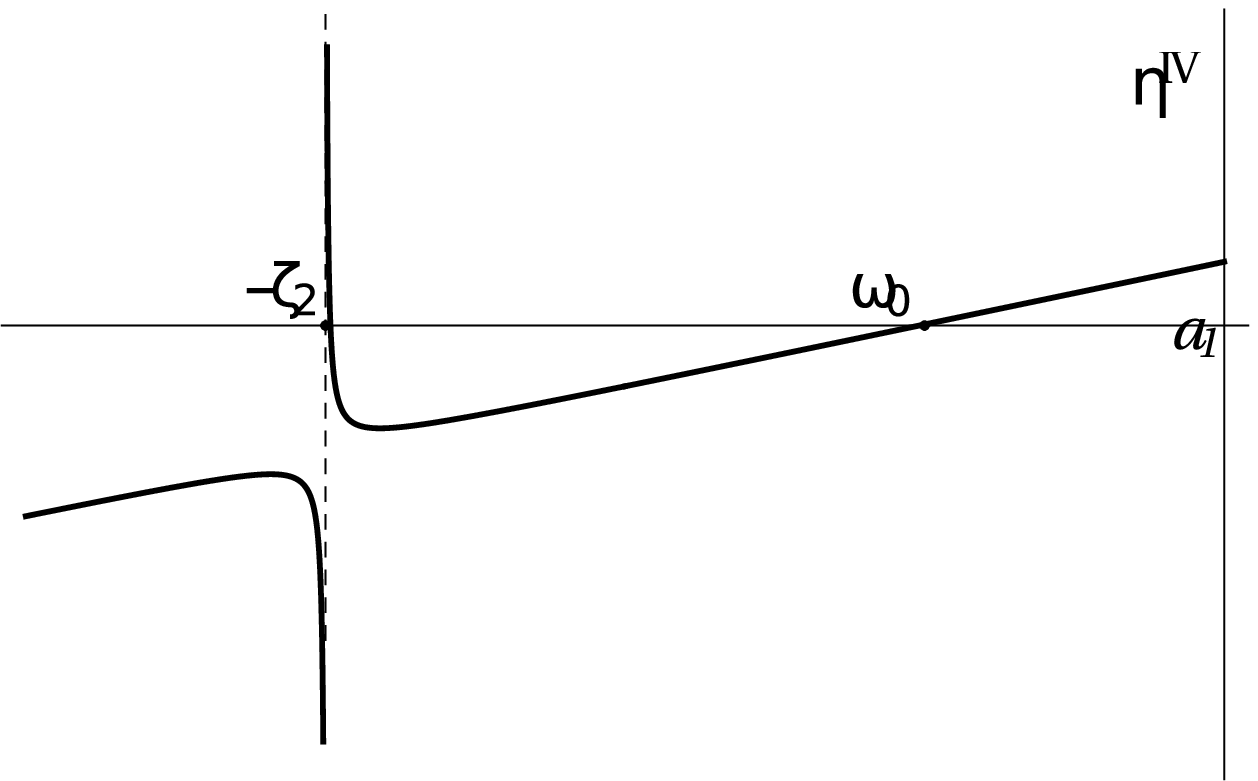}
\caption{$\eta(z)$ below the first threshold on different Riemann
sheets for $\zeta_1\neq\zeta_2\neq\omega_0<a_1$. \label{fig:eta-case1}}
\end{figure}

\begin{figure}
\includegraphics[height=2cm]{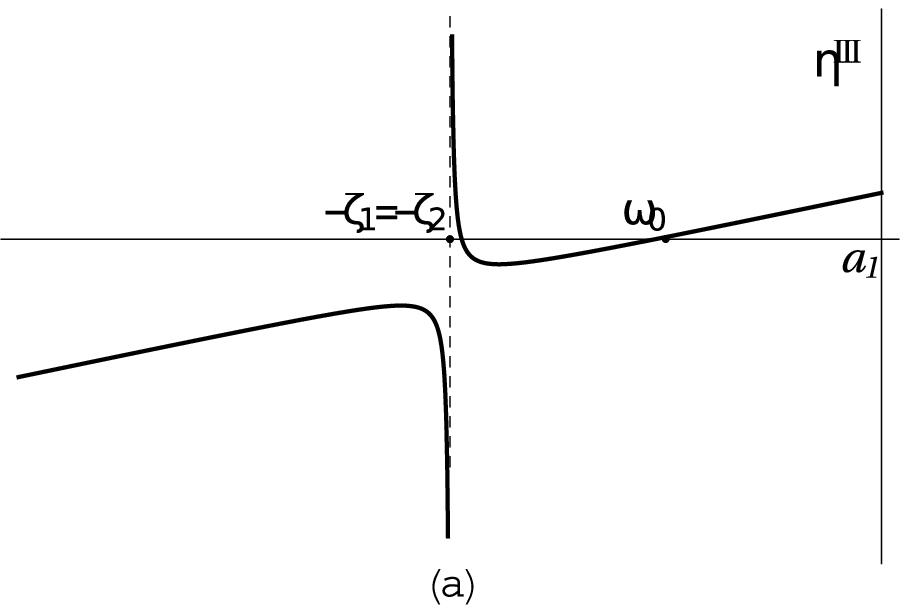}
\quad
\includegraphics[height=2cm]{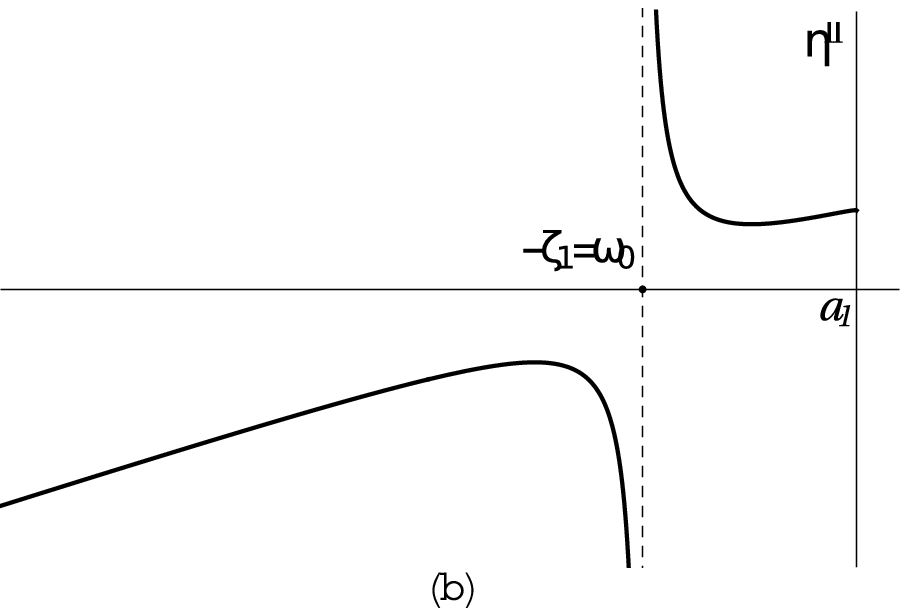}
\hspace{.5cm}
\includegraphics[height=2cm]{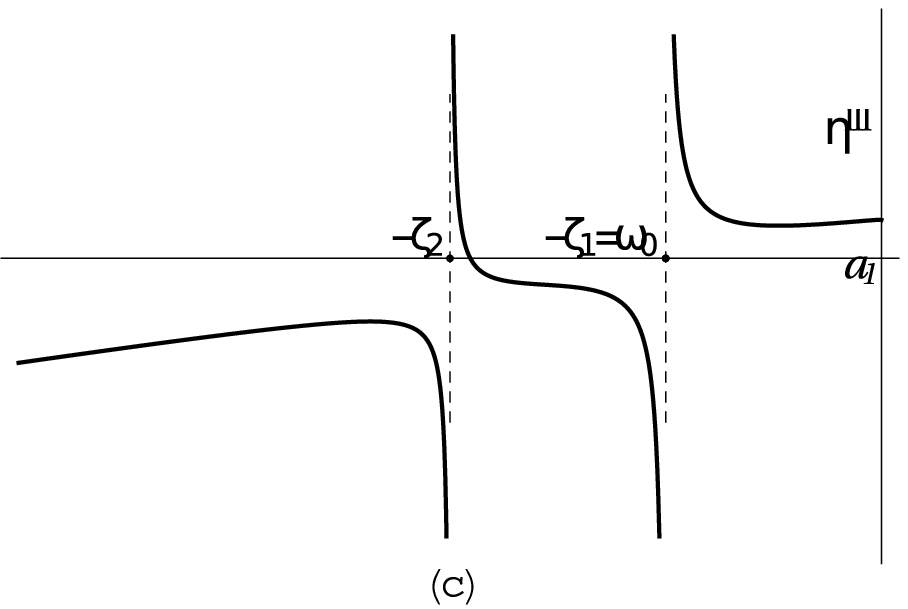}
\hspace{0.5cm}
\includegraphics[height=2cm]{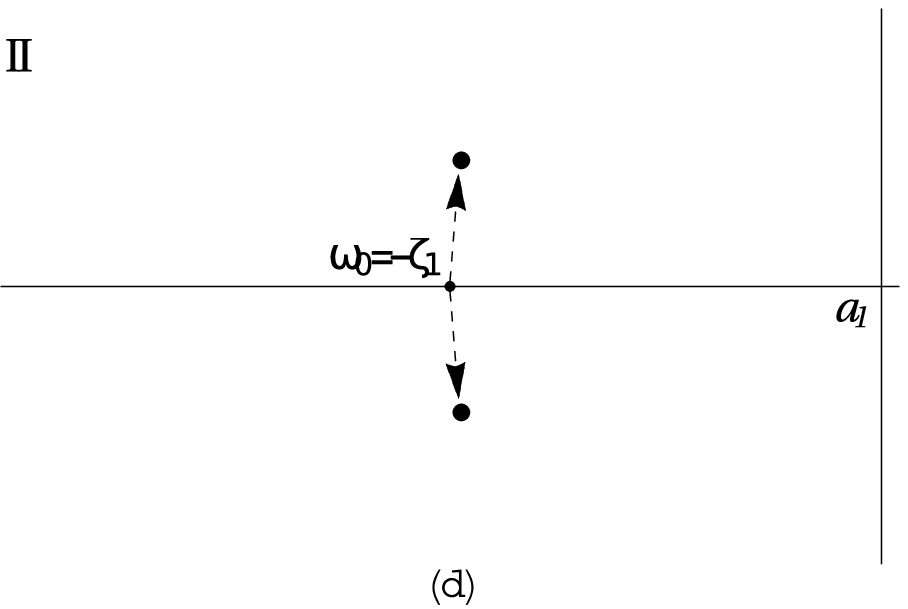}
\hspace{.5cm}
\includegraphics[height=2cm]{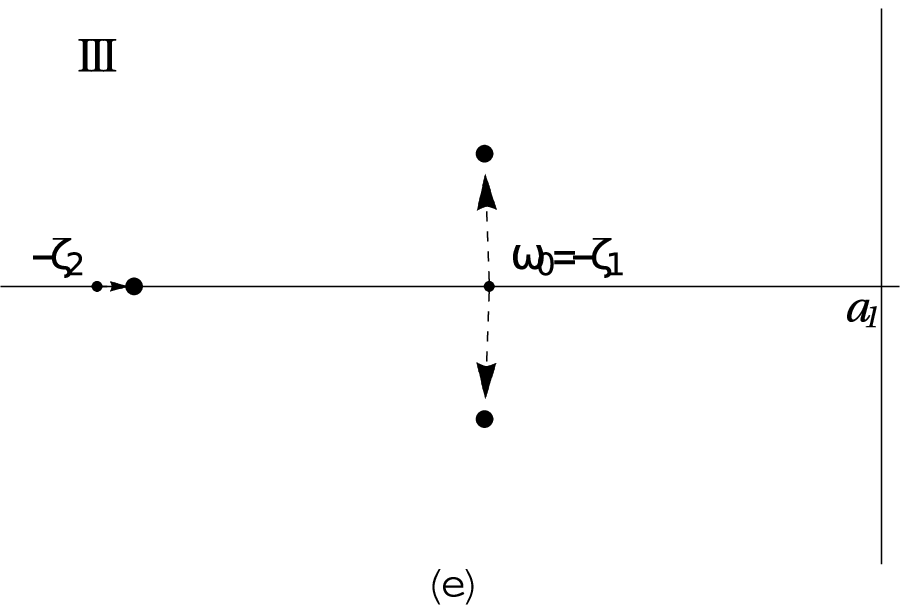}
\caption{(a): $\eta(z)$ on the third sheet below the first threshold for
$\zeta_1=\zeta_2\neq\omega_0<a_1$. (b) and (c): for
$\eta(z)$ on the second and third sheet for
$\zeta_2\neq\zeta_1=\omega_0<a_1$. (d) and (e): the pole positions
(large points) corresponding to
cases (b) and (c). Arrows denote the direction of motions of  the
poles as the couplings are increasing from zero.
\label{fig:case23}}
\end{figure}

Similar to the
argument in the single-continuum case in Ref.~\cite{Xiao:2016dsx}, near the pole of $G_1$, i.e. near
$-\zeta_1$ on the second sheet and the third sheet in our example, the form factor
will generate a zero point of $\eta(x)$ on each of these two sheets. The argument goes as follows: near
$-\zeta_1$, on the second
sheet and the third sheet
$\eta(\omega)=0$ takes the form of
\begin{align}
\frac{\lambda_1^2 c_1(\omega)}{\omega+\zeta_1} =\omega
-\omega_0+\lambda^2_2 c_2(\omega)\Rightarrow (
\omega -\omega_0+\lambda^2_2 c_2(\omega))(\omega+\zeta_1
)=\lambda_1^2 c_1(\omega),
\label{eq:virtual-argu}
\end{align}
where $c_1(\omega)$ and $c_2(\omega)$ are real functions on the real axis
below the lowest threshold and regular at $-\zeta_1$. At
$\lambda_{1,2}\to 0$ limit both $\omega_0$ and $-\zeta_1$ are the
solutions to the right equation and are first-order zero points,
though, at exact $\lambda_1=0$, $-\zeta_1$ is not the solution for the
left equation. When
$\lambda_{1,2}$ goes away from zero, $\eta(\omega)$ on the real axis will
still be real  below the threshold and be
deformed smoothly. So, the solutions originating from $-\zeta_1$
will not disappear and just moves away on the real axis. Similar argument can be
applied to $-\zeta_2$. Thus, each pole of the integral generated by the
form factor on each sheet will
introduce a virtual state near it for small couplings.  To sum up,
for small enough $\lambda_{1,2}$ and $\zeta_1\neq\zeta_2$, there are
virtual state poles generated
around $-\zeta_1$ on the second and the third sheets, and around $-\zeta_2$
on the third and fourth sheets. The typical shape of the $\eta(z)$ on
different sheet below the first threshold are shown in
Fig.~\ref{fig:eta-case1}. This argument applies to the general form
factors with  simple poles below the lowest threshold or with a pair
of simple poles on the complex plane.
If the form factors has multi-poles,
for example, a second-order pole below the threshold on the real axis, the poles on the
unphysical sheet may also move away onto the
complex plane becoming resonance poles by the similar argument. If in
some accidental cases, $\zeta_1=\zeta_2$, since  on the third sheet the
two poles of the integrals are combined
together, Eq.~(\ref{eq:virtual-argu}) becomes
\begin{align}
\frac{\lambda_1^2 c_1(\omega)+\lambda_2^2 c_2(\omega)}{\omega+\zeta_1} =\omega
-\omega_0\Rightarrow (
\omega -\omega_0)(\omega+\zeta_1
)=\lambda_1^2 c_1(\omega)+\lambda^2_2 c_2(\omega).
\label{eq:virtual-argu-III}
\end{align}
Thus, on the third sheet, there would be only one simple pole generated by the
two form factors. See Fig.~\ref{fig:case23}(a) for an illustration.

We then look at the poles generated from the original discrete state.
For $\omega_0<a_1$, at $\lambda_{1,2}=0$,
$\eta(\omega)=\omega-\omega_0 =0$  has the
solution $\omega=\omega_0$. Similar to the virtual state poles generated by
the form factors, as $\lambda_{1,2}$ are turned on, since near
$\omega_0$ both integrals in  $\eta(\omega)$ are real on each sheet,
$\eta(\omega)$ is only continuously corrected by a small real part  on the
negative axis on each sheet, and hence the original solution at $\omega_0$ is copied
on each sheet and moves away from $\omega_0$ on the real axis. So,
originating from the discrete state at $\omega_0$, there are one bound
state on the first sheet and one virtual state pole on each unphysical
sheet for small enough couplings. See Fig.~\ref{fig:eta-case1}  for an
illustration. Since these poles are below the
lowest threshold, only the bound state pole and the virtual state pole
on the second sheet are close to the physical region so that they may
have observable effects in experiments.

\begin{figure}
\includegraphics[height=2.5cm]{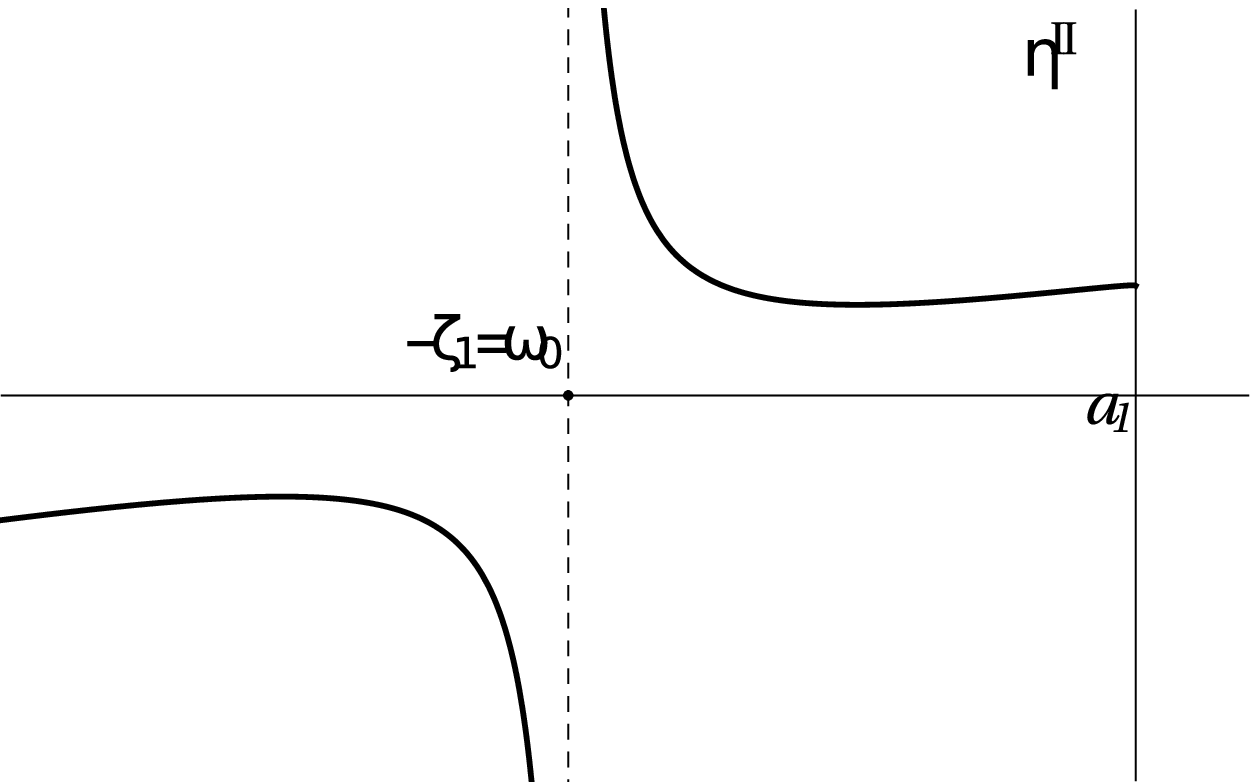}
\quad
\includegraphics[height=2.5cm]{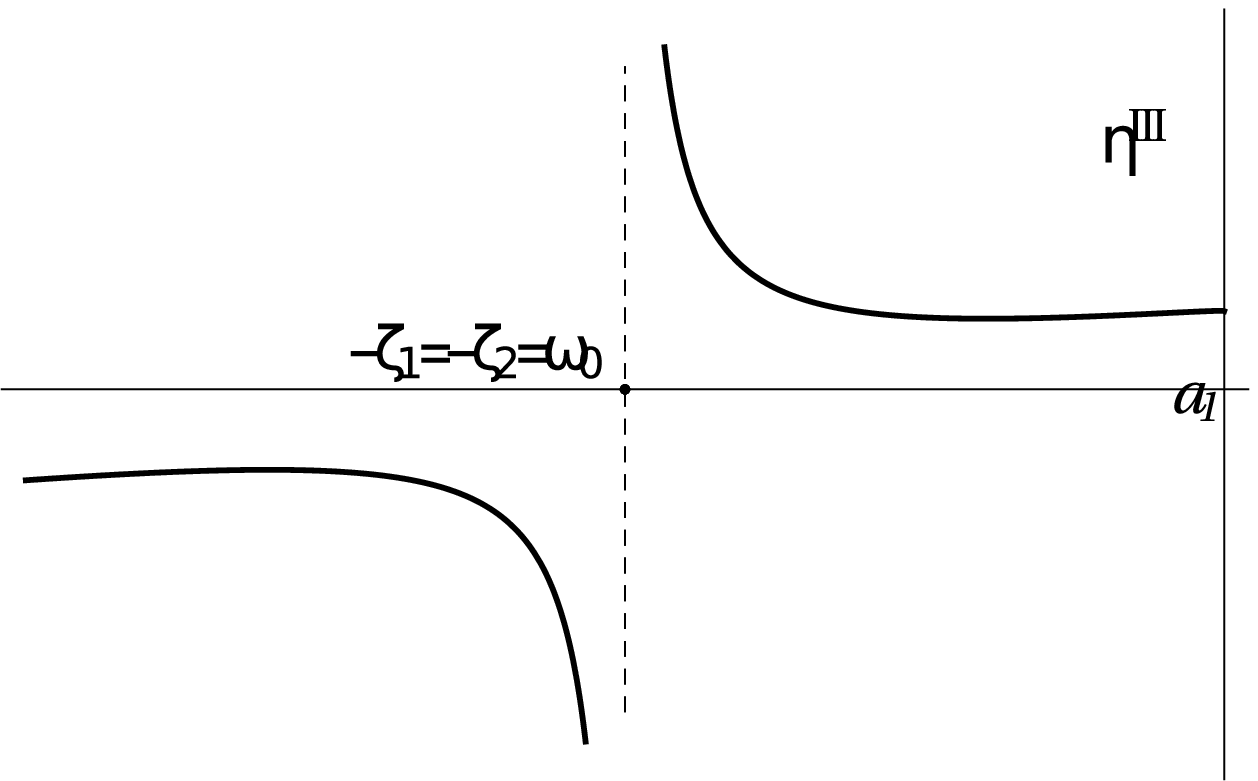}
\hspace{.5cm}
\includegraphics[height=2.5cm]{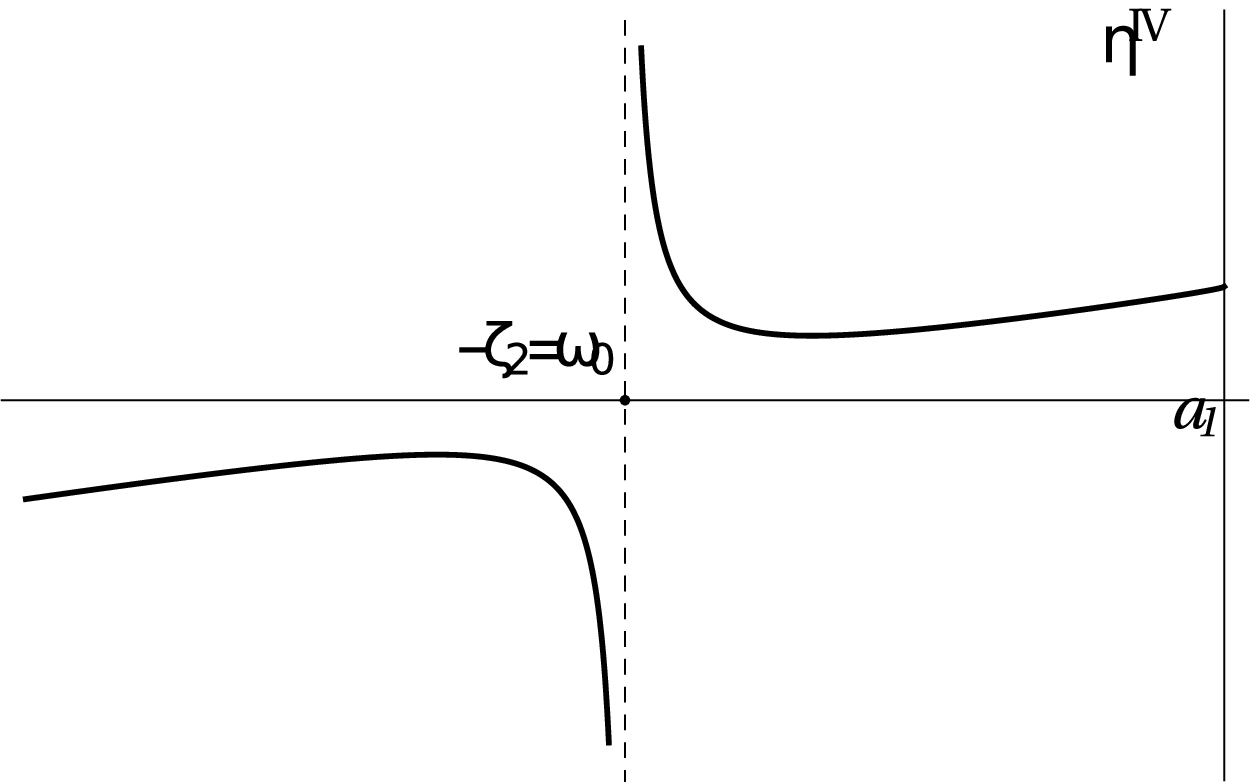}
%\hspace{0.5cm}
\\
\includegraphics[height=2.5cm]{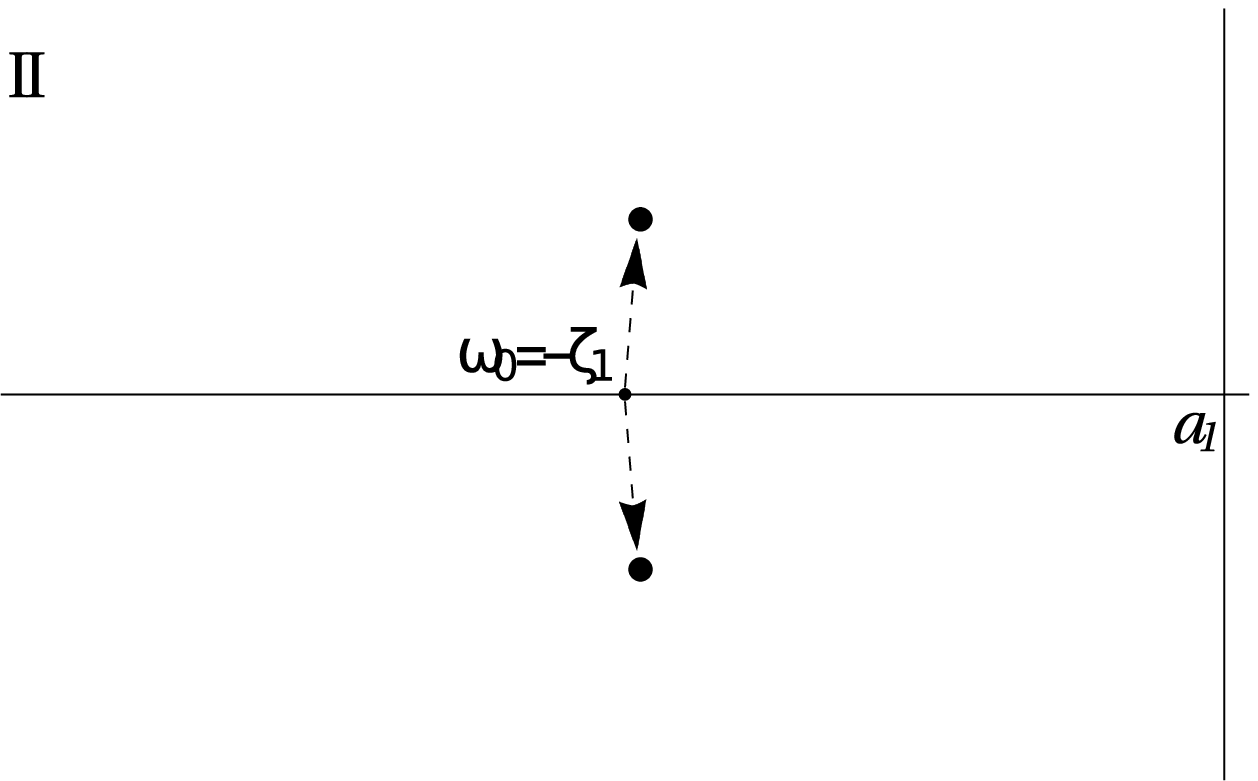}
\hspace{.5cm}
\includegraphics[height=2.5cm]{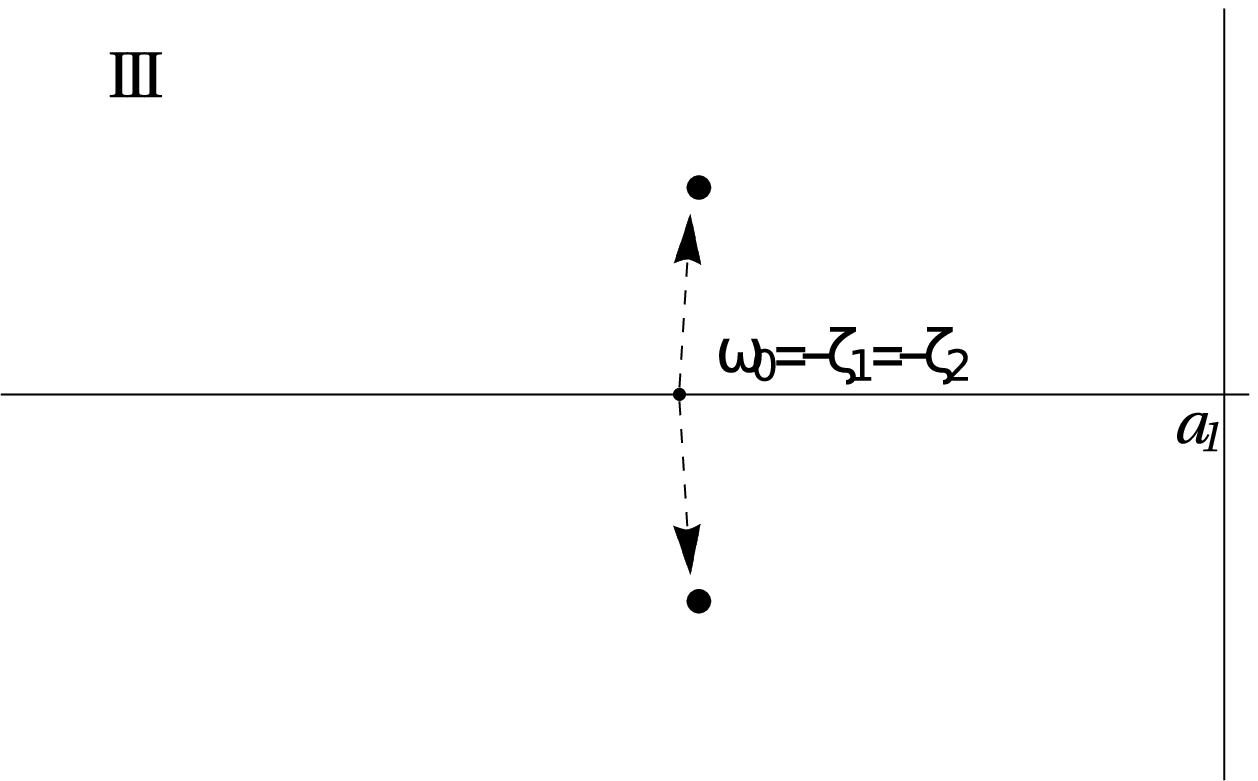}
\hspace{.5cm}
\includegraphics[height=2.5cm]{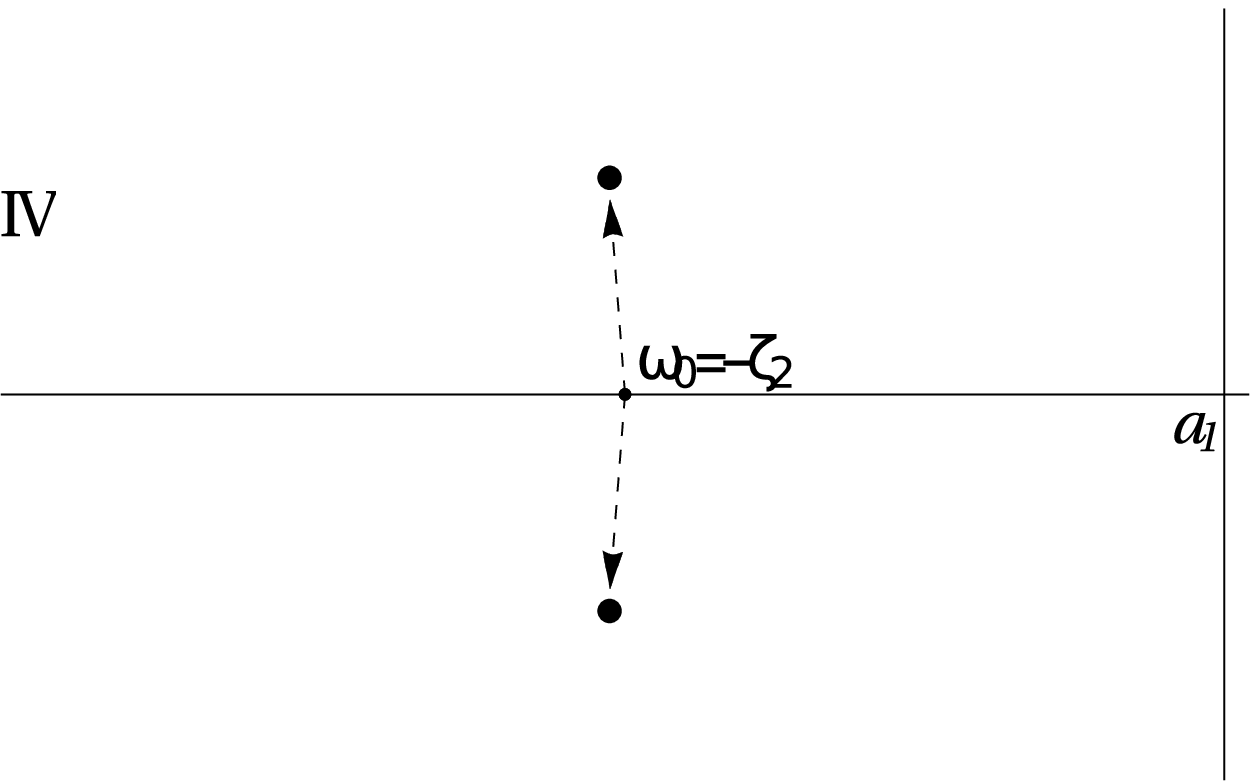}
\caption{$\eta(z)$ on the $II$, $III$, $IV$ sheet below the first threshold for
$\zeta_1=\zeta_2=\omega_0<a_1$ and the corresponding poles.
\label{fig:case4}}
\end{figure}

Concerning the virtual state poles from the form factors, there could be
another accidental case where $\omega_0=-\zeta_1$ and
$\zeta_1\neq\zeta_2$. In this case, Eq. (\ref{eq:virtual-argu}) can
also be used with
$\omega_0=-\zeta_1$. For the example form factors, it can be proved
that $c_1(\omega_0)<0$. If we turn on $\lambda_1$ first and then turn on
$\lambda_2$ slowly, the pole of $1/\eta$ generated by $\omega_0$ and
the pole generated near $-\zeta_1$ on each sheet where both appear would go into the complex plane and
become a pair of resonance poles symmetric with respect to the real
axis. See Fig.~\ref{fig:case23} for an
illustration. If for some form factors, $c_1(\omega_0)>0$, the two poles will
separate on the real axis. If we turn on $\lambda_2$ first and then turn on
$\lambda_1$, the two virtual state poles would separate first on the real
axis for small enough couplings. The other accidental case where $\omega_0=-\zeta_2$ and
$\zeta_1\neq\zeta_2$ can also be discussed similarly.
The most accidental case is that $\omega_0=-\zeta_1=-\zeta_2$
where Eq.(\ref{eq:virtual-argu-III}) can also be used with
$\omega_0=-\zeta_1$.
In this case, the two poles of the integrals from the form factors would combine
as discussed previously and on each unphysical sheet the two
solutions generated by
$\omega_0$ and the form factors would go into the complex plane and
become resonance poles if the right hand side of
Eq.(\ref{eq:virtual-argu-III}) is negative as for our example
form factors~(Fig.~\ref{fig:case4}), otherwise they will separate on the real
axis.

\begin{figure}
\includegraphics[height=3cm]{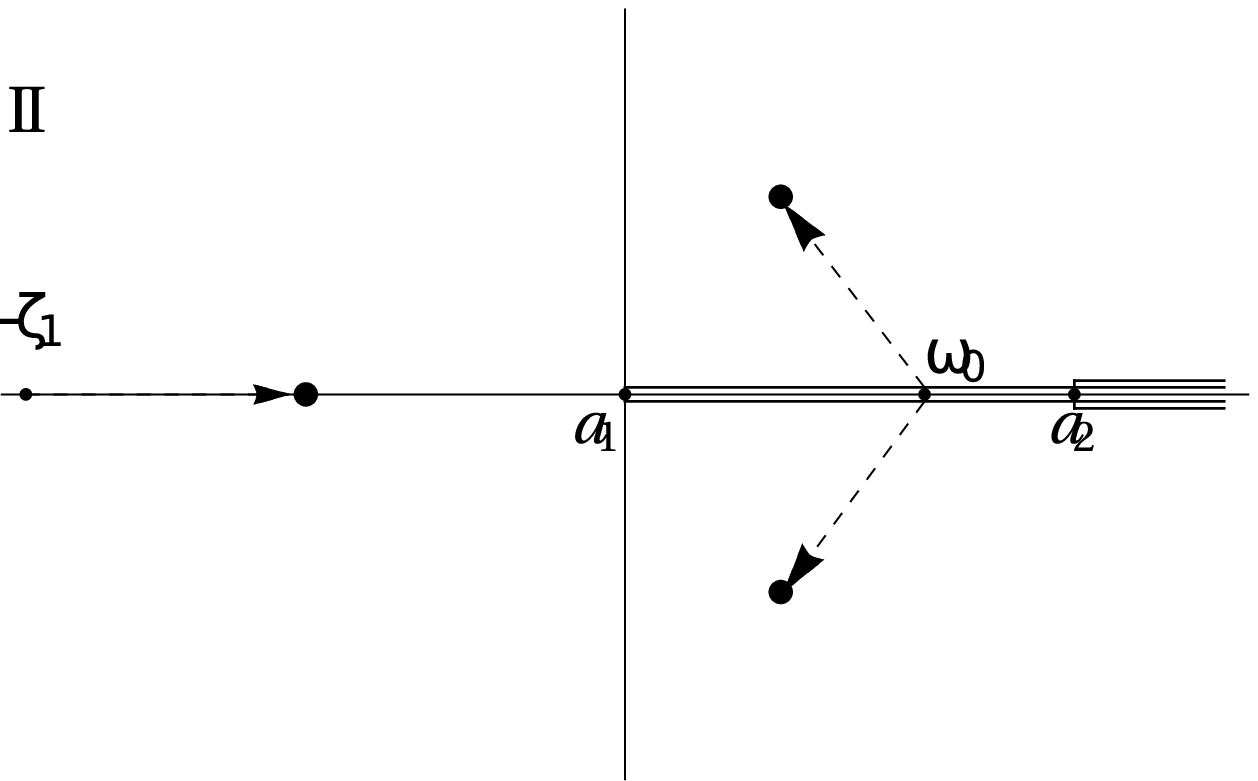}
\hspace{.5cm}
\includegraphics[height=3cm]{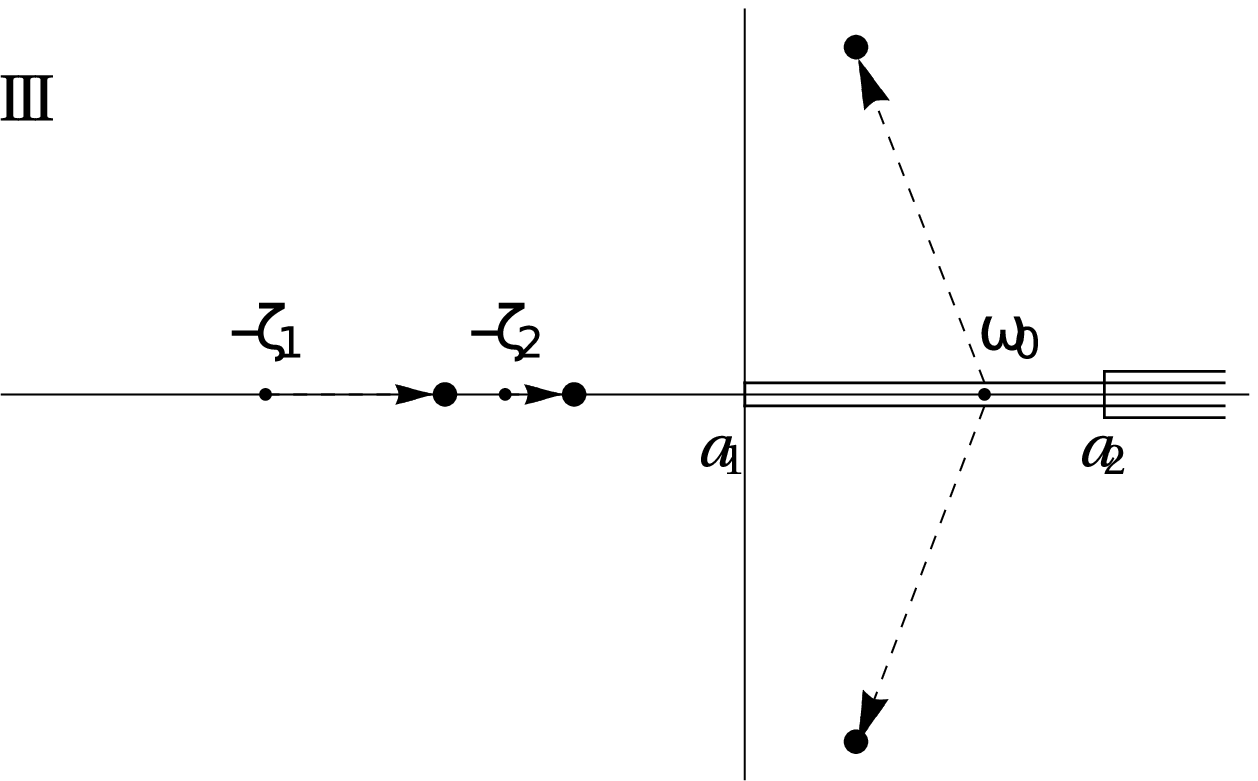}
\hspace{.5cm}
\includegraphics[height=3cm]{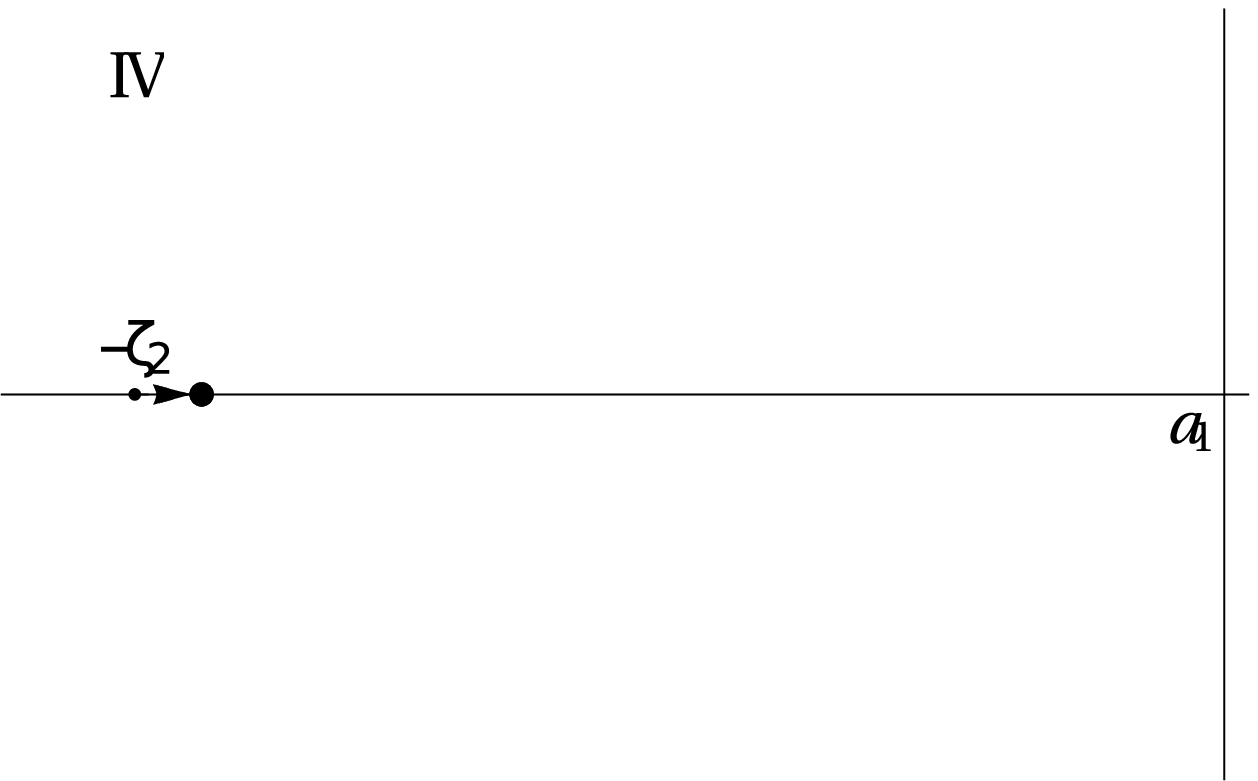}
\caption{Pole positions on  Riemann
sheets $II$, $III$, $IV$ for $\zeta_1\neq\zeta_2, a_1< \omega_0<a_2$.
The first sheet has no pole. \label{fig:case5}}
\end{figure}

For $a_1<\omega_0<a_2$, when $\lambda_1=0$, and $\lambda_2$ small
enough, we reduce to the single-continuum case, in which there is a
bound state on the first sheet and a virtual state on the second
sheet generated from $\omega_0$~\cite{Xiao:2016dsx}. Then when $\lambda_1$ is turned on,  the first sheet is duplicated to
be the first and the second sheet and the old second sheet is
duplicated to be the third and the fourth sheet.
The old bound state
on the first sheet will obtain finite decay width and move onto the
second sheet becoming a pair of resonance poles as shown in
Fig.~\ref{fig:case5}.  This can be
understood  as
follows. The analytically continued $\eta(x)$ on the first sheet and
the second sheet can be represented as
\begin{align}
\eta^{I}(x)=&x-\omega_0-\lambda_1^2\int_{a_1}^\infty\frac{G_1(\omega)}{x-\omega}\mathrm{d}\omega
-\lambda_2^2\int_{a_2}^\infty\frac{G_2(\omega)}{x-\omega}\mathrm{d}\omega
\,.
\\
\eta^{II}(x)=&x-\omega_0-\lambda_1^2\int_{a_1}^\infty\frac{G_1(\omega)}{x-\omega}\mathrm{d}\omega
-\lambda_2^2\int_{a_2}^\infty\frac{G_2(\omega)}{x-\omega}\mathrm{d}\omega
-2\pi i\lambda_1^2 G_1(x)\,.
\end{align}
 For
$x\sim \omega_0+O(\lambda^2_{1,2})$ on the upper plane (we take
$\lambda_1$ and $\lambda_2$ of the same order here), the first
integral term contributes a positive imaginary part to $\eta^I$,
$\sim \pi \lambda_1^2 G_1(x)+O(\lambda_{1,2}^4)$. The second integral
also gives a positive imaginary part $\sim O(\lambda_{1,2}^4)$.
Since the integral in the $\eta^I$ only has a positive imaginary
part near $\omega_0$ on the upper half plane, there could not be a
solution on the first sheet as the couplings are turned on.
 However, the last term in $\eta^{II}$ of the second sheet
 provides a  negative imaginary part around $\omega_0$. So, the solution to $\eta(x)=0$
must move from the original first sheet $\omega_0$ to the second sheet and for  real
analyticity, the solutions should be symmetric with respect of
the real axis. This is consistent with the causality. This argument
is independent of the specific form factors chosen.

Similarly, the original virtual state on the old second sheet when
only $\lambda_2$ is turned on will
move to the third sheet for the chosen example form factors, which can be seen as follows.  $\eta(\omega)$ on  the third and fourth sheet can be represented as
\begin{align}
 \eta^{III}(x)=&x-\omega_0-\lambda_1^2\int_{a_1}^\infty\frac{G_1(\omega)}{x-\omega}\mathrm{d}\omega
-\lambda_2^2\int_{a_2}^\infty\frac{G_2(\omega)}{x-\omega}\mathrm{d}\omega
-2\pi i\lambda_1^2 G_1(x)-2\pi i\lambda_2^2 G_2(x)
\\
\eta^{IV}(x)=&x-\omega_0-\lambda_1^2\int_{a_1}^\infty\frac{G_1(\omega)}{x-\omega}\mathrm{d}\omega
-\lambda_2^2\int_{a_2}^\infty\frac{G_2(\omega)}{x-\omega}\mathrm{d}\omega
-2\pi i\lambda_2^2 G_2(x)
\end{align}
For $x\sim \omega_0+O(\lambda^2_{1,2})$ with $O(\lambda_{1,2}^2)$
imaginary part on the upper plane, $-2\pi i
\lambda_2^2G_2(x)=-2\pi i \lambda_2^2\frac {\sqrt
{x-a_2}}{x+\zeta_2}$  will have $O(\lambda_{1,2}^4)$
imaginary part which can not cancel the $O(\lambda_{1,2}^2)$ imaginary
part of $x$. Only the $-2\pi i\lambda_1^2 G_1(x)$ on the third sheet
can cancel the imaginary part and  there can be solutions in the third
sheet complex plane as $\lambda_1$ is turned on. Thus, the pole will move onto the
third sheet. This argument also applies for general form factors.
This is because $\pi i\lambda_2^2 G_2$ gives the imaginary part of the second
integral, and should be real-analytic and has a cut starting from $a_2$.
Below the branch point $a_2$ of the cut, it is continuous and should be real by
real-analyticity. So, for
$x=\omega_0+iO( \lambda_{1,2}^2)$, the imaginary part of $-2\pi i
\lambda_2^2G_2(x)$ is always of $O(\lambda_{1,2}^4)$. Thus, the above
argument applies and in general the poles will move to the third sheet.

Alternatively, we could  turn off $\lambda_2$, and turn on
$\lambda_1$ first, and the discrete state will move to the second sheet to
be a pair of resonance poles in the reduced single-continuum case. Then when we turn on $\lambda_2$ slowly, the
first sheet will be doubled to become the first and the fourth sheet and
the second sheet will be duplicated to be the second sheet and the third
sheet with the resonance poles also copied to both sheets and
corrected by $O(\lambda_2^2)$. This argument  also applies for other general form factors
and the result is consistent with the one in the previous paragraph.
Turning on which coupling first may give
different trajectories though their final positions can be the same.

The last case is when the discrete state is above the second
threshold $\omega_0>a_2$. Similar to the above analysis, if
$\lambda_2=0$, and $\lambda_1$ is turned on, then the discrete state
will move to the second sheet becoming a pair of resonance poles and will
not appear on the first sheet. When we
turn on $\lambda_2$, the old first sheet will be duplicated to be the
first sheet and the fourth sheet. So, on these two
sheets there is no  pole originating from the
discrete state for small enough $\lambda_2$. The second sheet will be duplicated to be the second
sheet and the third sheet,  both carrying the resonance poles originated
from the discrete state. So, similar to above case, the discrete pole
will generate one pair of resonance poles on the second sheet and
another pair on the third sheet.

\begin{figure}
\includegraphics[height=3cm]{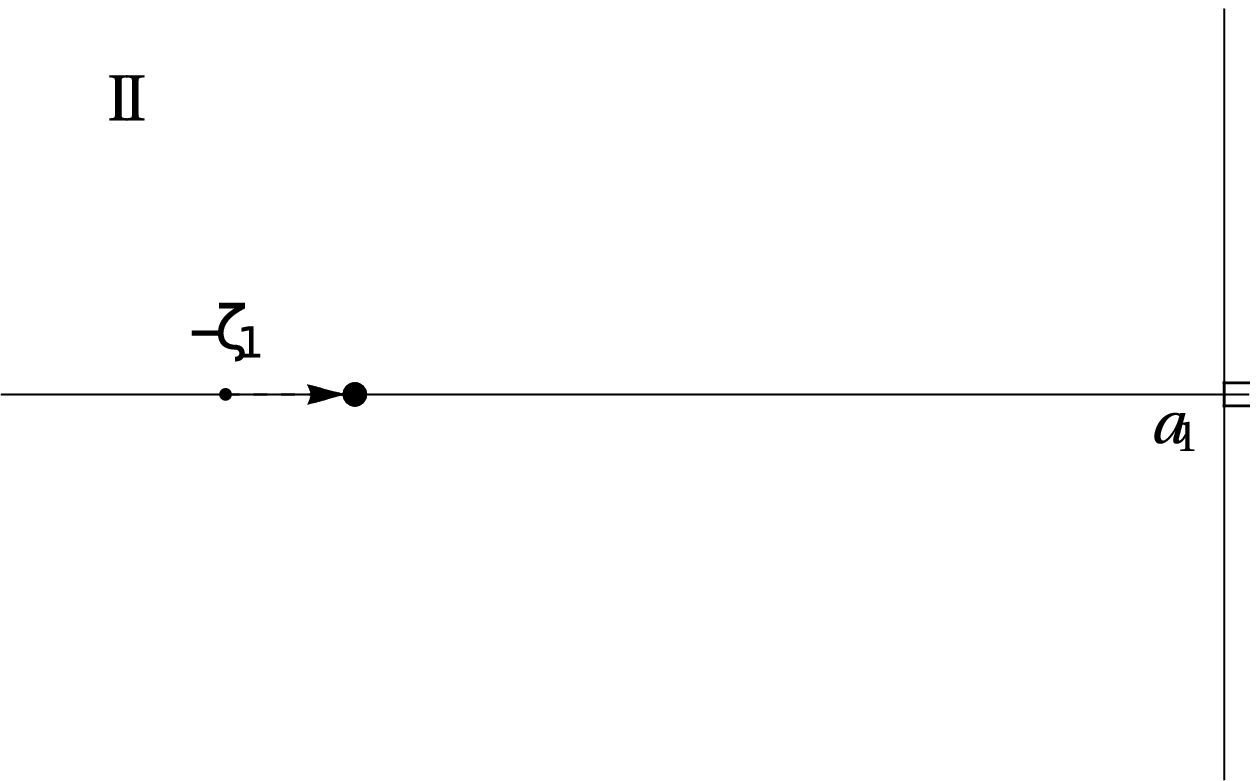}
\hspace{.5cm}
\includegraphics[height=3cm]{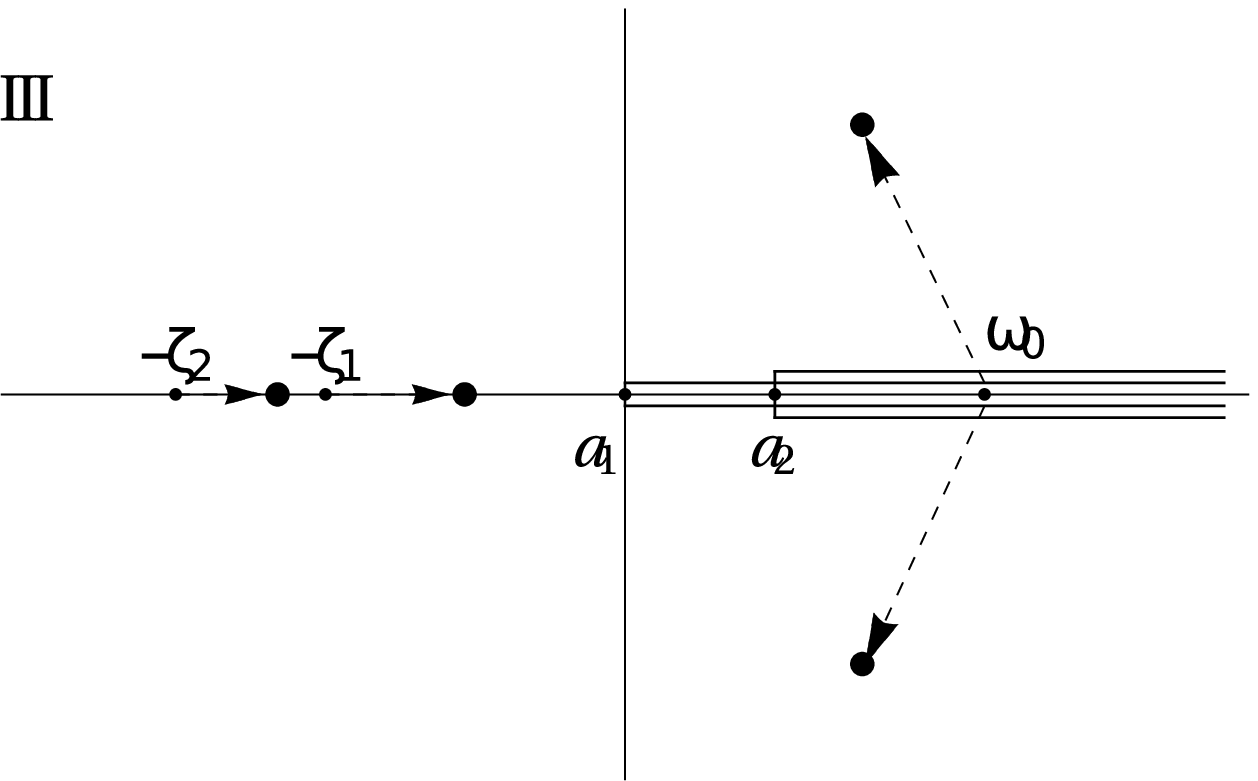}
\hspace{.5cm}
\includegraphics[height=3cm]{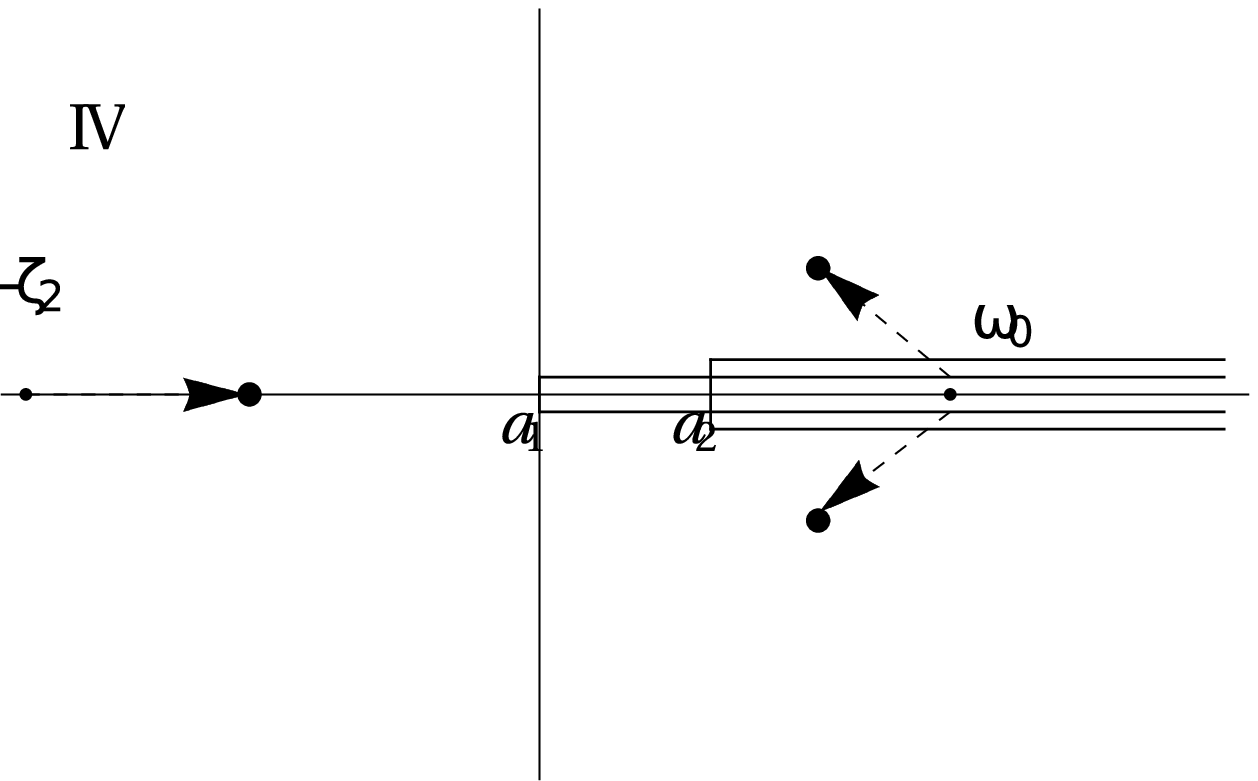}
\caption{Pole positions on  Riemann
sheets $II$, $III$, $IV$ for $\zeta_1\neq\zeta_2,  \omega_0>a_2$ when
turning on $\lambda_2$ first and then turning on small $\lambda_1$ slowly.
The first sheet has no pole. \label{fig:case6}}
\end{figure}

\begin{figure}
\includegraphics[height=3cm]{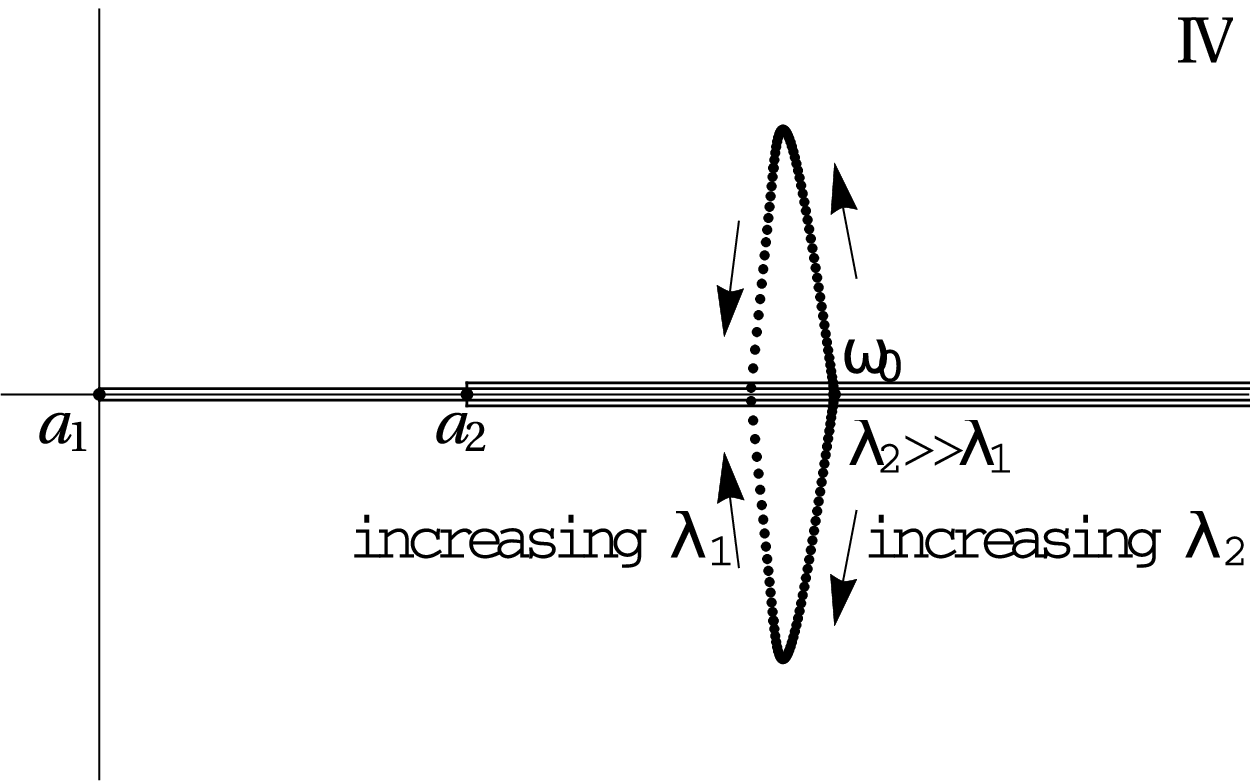}
\hspace{.5cm}
\includegraphics[height=3cm]{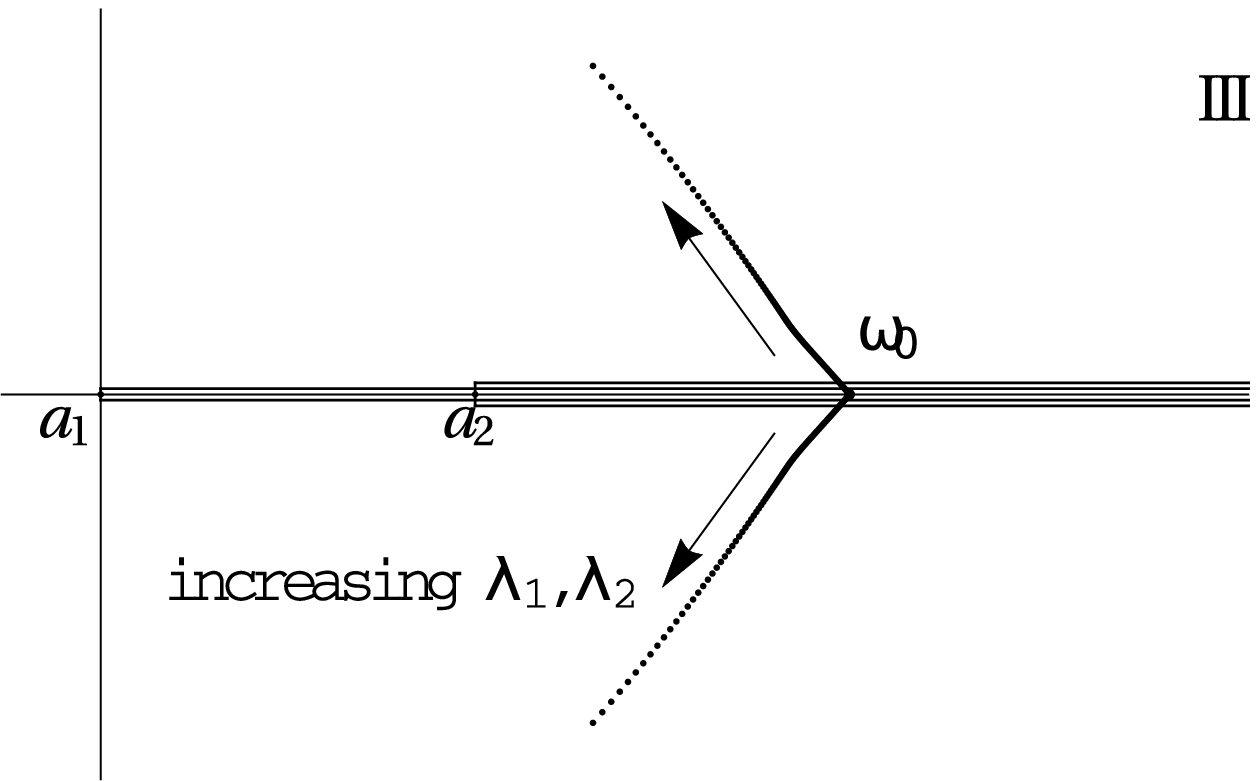}
\hspace{.5cm}
\includegraphics[height=3cm]{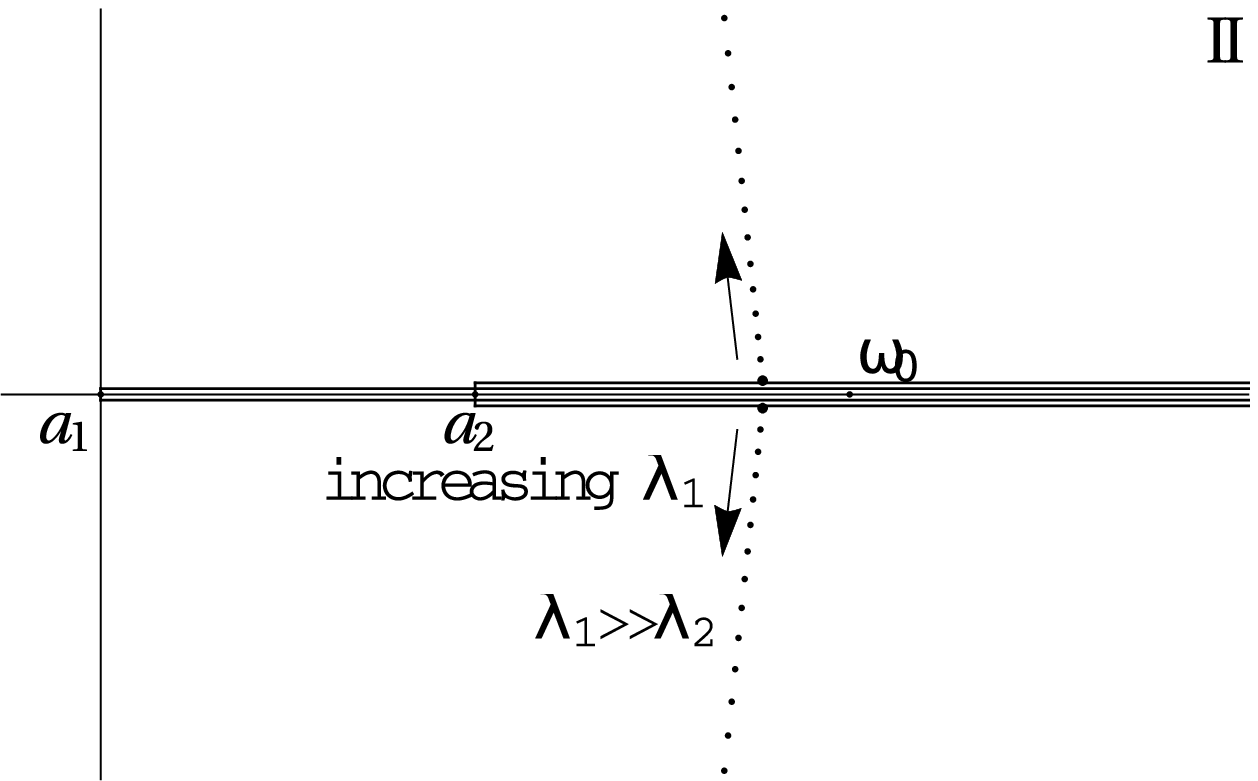}
\caption{When first turn on $\lambda_2$ and then turn on $\lambda_1$,
the discrete state moves to the $III$ and $IV$ sheets. After
increasing $\lambda_1$ to some extent, the $IV$ sheet poles run
across the second cut to the $II$ sheet. \label{fig:case7}}
\end{figure}

However, if we  first turn on $\lambda_2$, the discrete state will
become the resonance state on the second sheet. When we then turn on
$\lambda_1$ slowly, the old first sheet becomes the first and the second
sheets, and thus there are no poles moving onto these two sheets for
small enough $\lambda_1$. The old second
sheet becomes the third and the fourth sheet carrying the resonance
pole. Therefore, in this case the discrete state becomes the resonance
states on the third and the fourth sheets as shown in Fig.~\ref{fig:case6}.
As $\lambda_1$ becomes
larger to some extent, the fourth sheet pole may run across the second cut moving
to the second sheet as if we turn on $\lambda_1$ first and then turn
on $\lambda_2$. See Fig.~\ref{fig:case7}
for an illustration.

From the above discussions, we have observed that  when one bare discrete state is coupled to more than
one continuum states with different thresholds, there could be more than one pair of resonance
poles
generated from this discrete state. We have two  thresholds here and
the number of the Riemann sheets is four which doubles the ones when
there is only one continuum. The
poles on the Riemann sheets will also be duplicated and thus the number of
the poles will also be doubled.

\section{Discrete states and the completeness relation\label{sect:Gamow}}

We have seen that there could also be poles on the second, third and fourth
sheets.
One can prove that the bound state $|\Psi_0(x_B)\rangle$ if
 exists, and
these two continuum states form a
set of
complete basis,
\begin{align}
|\Psi_0(x_B)\rangle\langle\Psi_0(x_B)|+&\int_{a_1}^\infty dx\, |\Psi_{1\pm}(x)\rangle\langle
\Psi_{1\pm}(x)|+\int_{a_2}^\infty dx\, |\Psi_{2\pm}(x)\rangle\langle
\Psi_{2\pm}(x)|
\nonumber\\=&
|1\rangle\langle 1| +\int_{a_1} d\omega
|\omega\rangle_1{ }_1\langle\omega|+\int_{a_2} d\omega
|\omega\rangle_2{ }_2\langle \omega|=\mathbf 1\,.
\end{align}
However the other discrete states on unphysical sheets do not enter the completeness
relation.

\begin{figure}
\includegraphics[width =5cm]{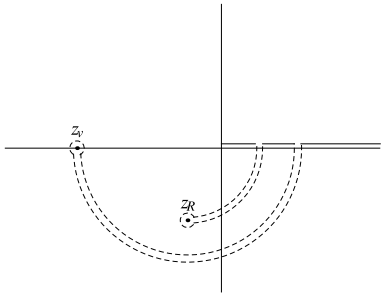}
\caption{A typical integral path used in define the continuum ket states in
the single-continuum case. The dashed line denotes the deformation of
the contour into  the second sheet.\label{fig:contour-V-R}}
\end{figure}

In the simplest Friedrichs model,  in Ref.~\cite{Prigogine:1991}, in order to
solve the large Poincar\'e problem and based on two physical
conditions: the decay of unstable state in the future and the emission
of out-going wave, PPT take the continuum states
$|\Psi_{\pm}(x)\rangle$ as a complex functional and choose a kind of the
integral contour for $x$ in the continuum states, and then all the
discrete states enter the completeness equation equally. In their
discussion, the original discrete state is defined to be a little
above the real axis, and the integral path on the real axis is below the
discrete state. After the coupling is turned on, the discrete state
goes below the real axis to the second sheet complex plane, and the
integral contour should also be deformed to the lower half plane, below
the pole of the discrete state.
In Ref.~\cite{Xiao:2016dsx}, we also found that all the second sheet
poles could merge and separate, and these poles should not be treated
differently, whether they are generated from the discrete state or
from the form factor. The integral path for the continuum state should be deformed to the second sheet
around all the poles on the lower half of the second sheet like the
one in Fig.~\ref{fig:contour-V-R}.  After this modification to the
continuum state, all the discrete states including the ones on the second sheet
  enter the completeness relation.  Here, we can
generalize this kind of definition to the two-continuum case.

\begin{figure}
\includegraphics[width=5cm]{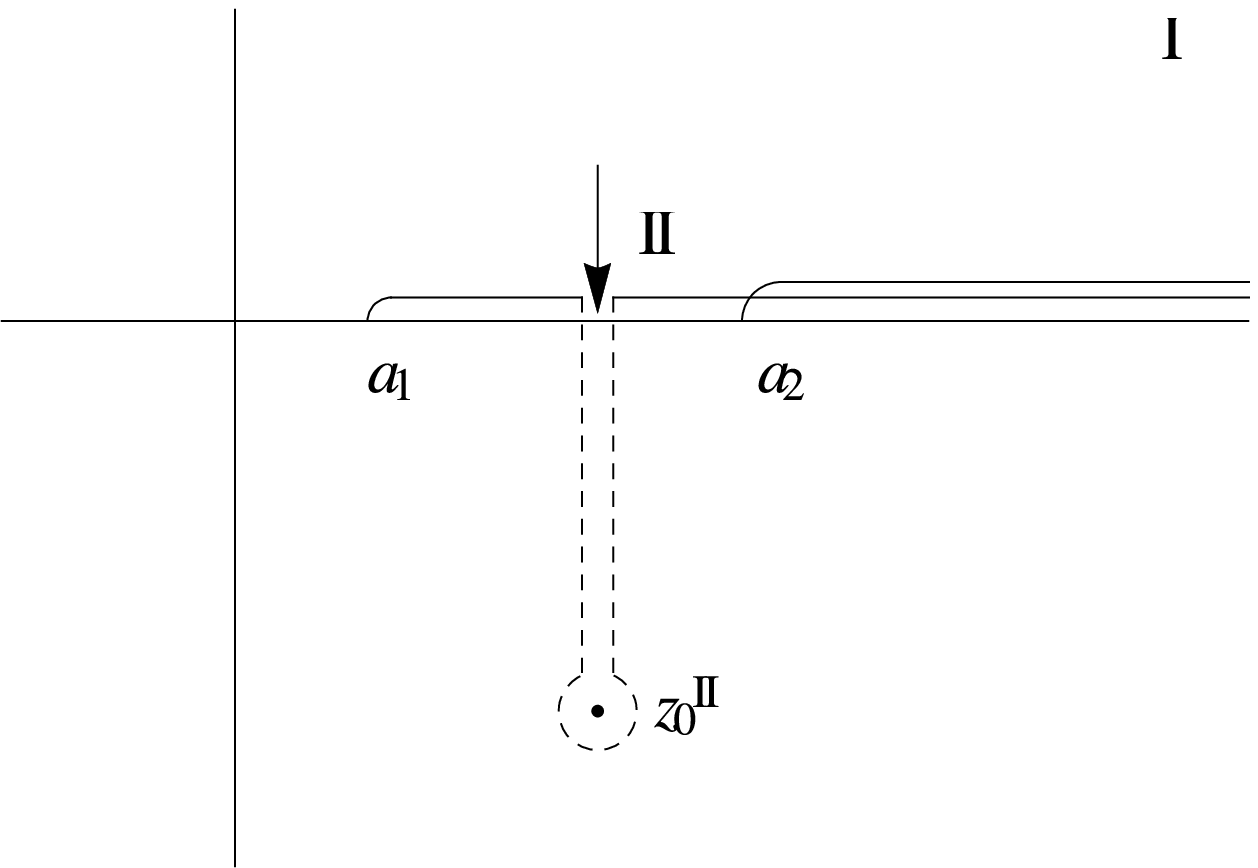}
\includegraphics[width=5cm]{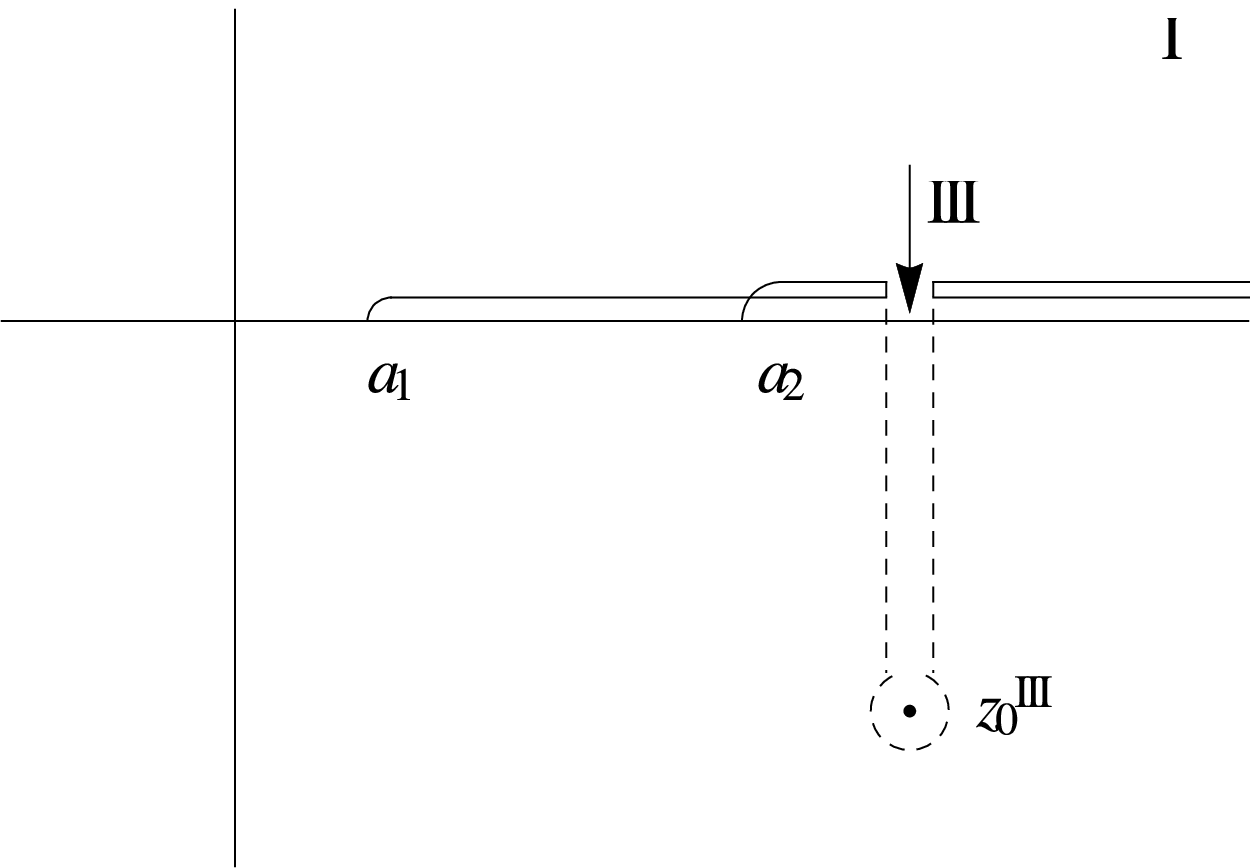}
\includegraphics[width=5cm]{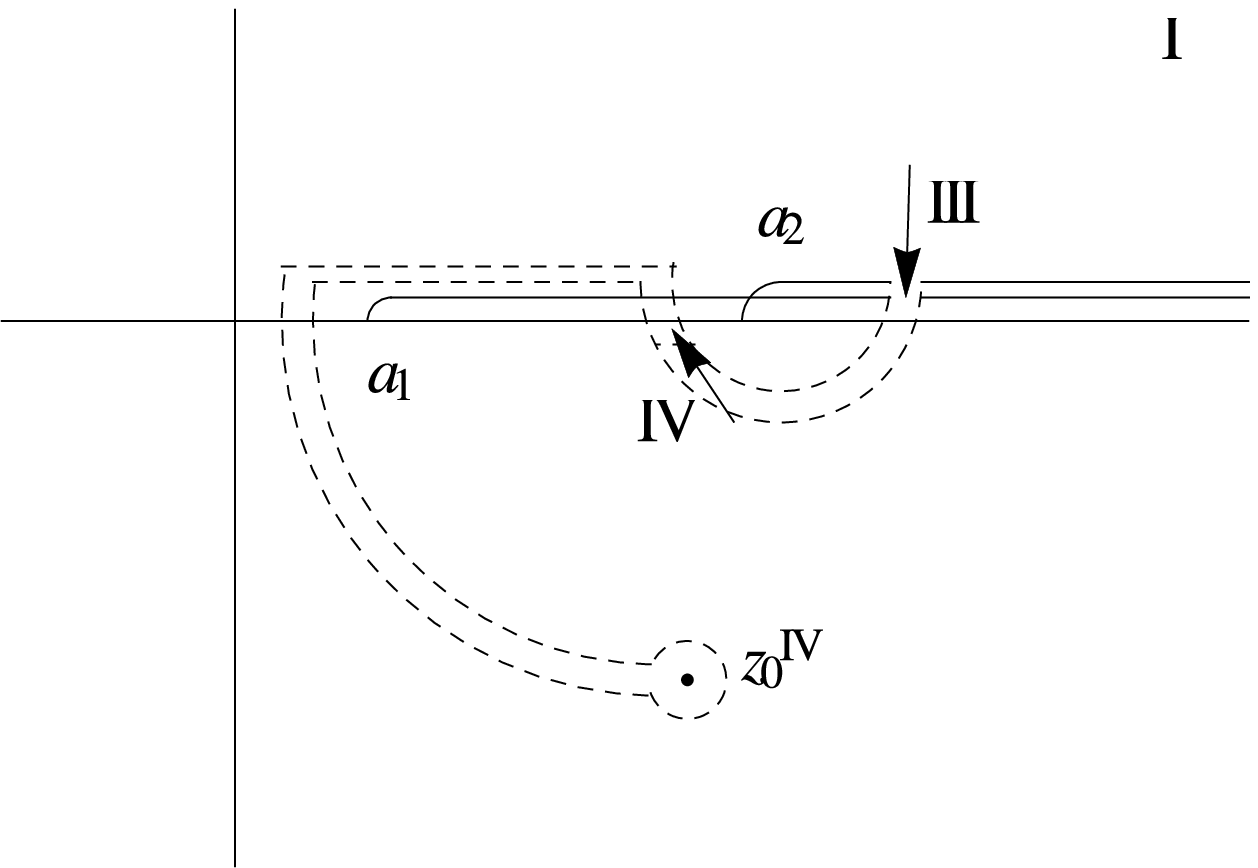}
\caption{Integral paths for $\eta_d^+$. $I$ means the first sheet.
 The $II$, $III$, $IV$ beside
the arrow denote the sheets to be analytically continued to.\label{fig:contour-c}
}
\end{figure}

In this spirit, we need to choose the integral contour for  $x$ in the $1/\eta^\pm(x)$
in
$|\Psi(x)\rangle$ for coupled channel cases. If we turn off one
of the couplings, we come back to the single-continuum case and the
contour should be the same as in the single-continuum case. So, the
integral contour should go from the first sheet to each sheet around
all poles and back to the
first sheet along the positive real axis to the
infinity as shown in Fig.~\ref{fig:contour-c}.  To be specific, if there are $N_{II}$ ($N_{III}$, $N_{IV}$) second
(third , fourth) sheet poles, including the poles generated from the
form factors, similar to \cite{Prigogine:1991}, the $\eta^{\pm}(\omega)$ in the
continuum right eigenstate in Eqs.
(\ref{eq:solu-final-1}) and (\ref{eq:solu-final-2}) should be modified
to
\begin{align}
\eta_d^{\pm}(\omega)\equiv\eta^{\pm}(\omega)\prod_{J=II,III,IV}\prod_{i=1}^{N_{J}}\frac{\omega-z_{i}^{J}}{[\omega-z_{i}^{J}]_{\pm}}\,.
\label{eq:etad-c}
\end{align}
This equation just means that we put the integral path information in the
definition of $\eta^{\pm}_d$ through $[\omega-z_i^J]_{\pm}$, where  $[\dots]_{+}$
means the integral path of $\omega$ is continued  as shown in Fig.
\ref{fig:contour-c}, and $[\dots]_-$ means just the
opposite.

 In the left eigenstates
the definition of the integral path on the real axis is not modified,
since there is no pole for $\eta^-$ continued downward from the lower
edge of the cut on the first sheet to the lower
half plane.
As in \cite{Prigogine:1991}, this kind of choice of the contour is
consistent with two physical conditions,  i.e., the decay of unstable state in the
future and the emission of out-going wave.
So the right and left continuum eigenstates can be expressed as
\begin{align}
|\Psi^d_{i\pm}(x)\rangle=|x\rangle_i+\frac{\lambda_i
f_i^*(x)}{\eta_d^\pm(x)}\Big[|1\rangle+\sum_{j=1,2}\lambda_j\int_{a_j}^\infty \mathrm
d\omega \frac{f_j(\omega)}{x-\omega\pm
i0}|\omega\rangle_j\Big]
\label{eq:solu-final-r-i}
\\
\langle \tilde \Psi_{i\pm}(x)|={ }_i\langle x|+\frac{\lambda_i
f_i(x)}{\eta^\mp(x)}\Big[\langle 1|+\sum_{j=1,2}\lambda_j\int_{a_j}^\infty \mathrm
d\omega \frac{f^*_j(\omega)}{x-\omega\mp
i0}{\ }_j\langle\omega|\Big]
\label{eq:solu-final-l-i}
\end{align}
and orthogonal relations are not modified, $\langle \tilde
\Psi_{i\pm}(x)|\Psi^d_{j\pm}\rangle=\delta_{ij}\delta(x-y)$.

We have shown that  there will be more than one Gamow states
corresponding to one original discrete state and could also
be other states generated from the
form factors. Since the poles can lie on the unphysical sheets, the
integral in the expression of the states need to be analytically
continued to different sheets and the integral path should
be deformed accordingly, that is,
\begin{align}
|z_0^{I}\rangle=&N^I\Big(|1\rangle+\lambda_1\int_{a_1}^\infty \mathrm
d\omega \frac{f_1(\omega)}{z^I_0-\omega}|\omega\rangle_1+\lambda_2\int_{a_2}^\infty \mathrm d\omega
\frac{f_2(\omega)}{z^I_0-\omega}|\omega\rangle_2\Big) \,,
\\
|z_0^{II}\rangle=&N^{II}\Big(|1\rangle+\lambda_1\int_{a_1}^\infty \mathrm
d\omega \frac{f_1(\omega)}{[z^{II}_0-\omega]_+}|\omega\rangle_1+\lambda_2\int_{a_2}^\infty \mathrm d\omega
\frac{f_2(\omega)}{z^{II}_0-\omega}|\omega\rangle_2\Big) \,,
\\
|z_0^{III}\rangle=&N^{III}\Big(|1\rangle+\lambda_1\int_{a_1}^\infty \mathrm
d\omega \frac{f_1(\omega)}{[z^{III}_0-\omega]_+}|\omega\rangle_1+\lambda_2\int_{a_2}^\infty \mathrm d\omega
\frac{f_2(\omega)}{[z^{III}_0-\omega]_+}|\omega\rangle_2\Big) \,,
\\
|z_0^{IV}\rangle=&N^{IV}\Big(|1\rangle+\lambda_1\int_{a_1}^\infty \mathrm
d\omega \frac{f_1(\omega)}{z^{IV}_0-\omega}|\omega\rangle_1+\lambda_2\int_{a_2}^\infty \mathrm d\omega
\frac{f_2(\omega)}{[z^{IV}_0-\omega]_+}|\omega\rangle_2\Big) \,,
\end{align}
which are the generalized eigenstates of $H$ with eigenvalue $z_0^J$
on the lower half plane of the $J$-th sheet with $J=I,II,III,IV$, including the real axis
below the lowest thresholds.
Since the two integrals defines the two cuts independently, the
integral path can be defines separately for each integral. The $[\dots]_+$ means the continuation from
upper half of the Riemann sheet to the pole positions on the
other Riemann sheets, through the first cut in the first integral or
through the second cut in the second integral, which requires the integral
paths to be deformed accordingly.
For example, the third sheet is reached by going through both  cuts above the second threshold; therefore
both integral paths are deformed.
To simplify notation, we encode the contour information into the pole
position, denoted by $z^J_{0\mp}$, for $J=I,II,III,IV$, in which the
subscript ``$-$" means continuation from upper to lower sheets and ``$+$"
means the opposite, and from $J$ one can
determine whether the continuation across the cut is needed or not.
One can then unify the states in the four sheets to be
\begin{align}
|z_0^{J}\rangle=&N^J\Big(|1\rangle+\lambda_1\int_{a_1}^\infty \mathrm
d\omega \frac{f_1(\omega)}{z^J_{0-}-\omega}|\omega\rangle_1+\lambda_2\int_{a_2}^\infty \mathrm d\omega
\frac{f_2(\omega)}{z^J_{0-}-\omega}|\omega\rangle_2\Big) \,,
\end{align}
For example, if $J=II$, only the first integral is continued downward and the
second integral is not.
It is easy to check directly that $H|z^J_0\rangle
=z^J_0|z^J_0\rangle$.
The left eigenstates with the same eigenvalue as the corresponding
right eigenstates can also be solved,
\begin{align}
\langle  \tilde z_0^{J}|=&N^J\Big(\langle1|+\lambda_1\int_{a_1}^\infty \mathrm
d\omega \frac{f_1^*(\omega)}{z^J_{0-}-\omega}{\ }_1\langle\omega|+\lambda_2\int_{a_2}^\infty \mathrm d\omega
\frac{f_2^*(\omega)}{z^J_{0-}-\omega}{\ }_2\langle\omega|\Big) \,,
\end{align}
The normalization constants are chosen to be
$N^J=1/(\eta'(z_0^J))^{1/2}$
such that $\langle\tilde z_0^{J}|z_0^{J}\rangle=1$ for $J=I,II,III,IV$.
These states on different sheets can be proved to be orthogonal to
each other, $\langle\tilde z_0^{L}|  z_0^{K}\rangle=0$ for $L\neq K$.

As we stated, in the  case with one continuum, after including the integral path into the continuum
states, the resonance poles and the virtual state
poles also enter the completeness relations \cite{Prigogine:1991,
Xiao:2016dsx}.
In two-continuum case, one discrete state is split onto
different Riemann sheets to be different resonant states or virtual
states. There are also other virtual states or resonances
generated by the form factors.
It is expected that all these discrete states enter into the
completeness relation using the continuum states defined above.
This can be proved as expected
\begin{align}
\sum_{i=1,2}\int_{a_i}^\infty dx|\Psi^d_i(x)\rangle\langle \tilde
\Psi_i(x)|+\sum_{J,i} |z_{0,i}^J\rangle \langle \tilde z_{0,i}^J|= \mathbf 1
\end{align}
where $i=1,\dots,N_J$, for $N_J$ poles on the $J$-th sheet.
We consider only simple poles here. In \cite{Xiao:2016dsx}, the
poles are found to merge to be higher order poles and separate for larger
couplings. Similar things could also happen here, and the same discussion in
the one-continuum case also applies which will not be repeated here.

\section{More than two continuum states\label{sect:general}}

We have discussed the solutions for Friedrichs model with two continuum
states. It is straightforward to extend the solution to the model with more than two
 continua. Suppose that there are $N$ continuum states
$|\omega\rangle_i$
coupled to a discrete state $|1\rangle$ without the direct couplings
among the continua, described by the Hamiltonian
\begin{eqnarray}
H&=&\omega_0|1\rangle\langle1|+\sum_{i=1}^N\int_{a_i}^\infty
\mathrm{d}\omega\omega|\omega\rangle_i{}_i\langle\omega|\nonumber\\
&+&\sum_{i=1}^N \lambda_i\int_{a_i}^\infty\mathrm{d}\omega[f_i(\omega)|\omega\rangle_i\langle
1|+f_i^*(\omega)|1\rangle{\ }_i\langle\omega|]\,.
\end{eqnarray}
The continuum state solutions for the full Hamiltonian can be obtained
similar to Eqs. (\ref{eq:solu-final-1}) and (\ref{eq:solu-final-2}):
\begin{align}
|\Psi_{i\pm}(x)\rangle=|x\rangle_i+\frac{\lambda_i
f_i^*(x)}{\eta^\pm(x)}\Big[|1\rangle+\sum_{j=1}^N\lambda_j\int_{a_j}^\infty \mathrm
d\omega \frac{f_j(\omega)}{x-\omega\pm
i0}|\omega\rangle_j\Big]\,,
\label{eq:solu-final-r-n}
\end{align}
where
\begin{align}
\eta^\pm(x)=x-\omega_0-\sum_{i=1}^N \lambda_i^2\int_{a_i}^\infty {\rm
d\omega}\frac
{G_i(\omega)}{x-\omega\pm i0}  ,\quad
G_i=f_i(\omega)f_i^*(\omega).
\end{align}
Whenever there is an additional continuum included, on each Riemann
sheet there is a new cut and analytically continuation will double the
Riemann sheets. Thus there are $2^N$ Riemann sheets.
The discrete states also correspond to the zero points of the
analytically continued $\eta(x)$
and will be carried to the duplicated Riemann sheets and be
renormalized separately. Thus, the number of the poles generated by the original
discrete states will also be $2^N$.
The poles introduced by the form factors will also  be
copied on the duplicated Riemann sheets when the other continua
channels are included, and hence the number for each such poles will
be $2^{N-1}$ except some accidental cases as discussed in Sect.
\ref{sect:polepos}. All the previous discussions can be extended to here
without any difficulty. We will not repeat it here. Although there
are so many poles on the unphysical sheets, only those near the
physical region may have observable effects in the experiments.

\section{Conclusion and discussions\label{sect:conclude}}

In this paper we have first discussed the solution for  the Friedrichs model
with one discrete state coupled to two continuum states with different
thresholds. The
generalized eigenvalues for the full Hamiltonian include the
continuum spectra and all the zero points for the $\eta(x)$ on the
four-sheeted Riemann surface. The generalized eigenstates for the continuum spectra
are expressed in Eq. (\ref{eq:solu-final-1}) and Eq. (\ref{eq:solu-final-2}) which reduce to the original two continuum
states when the couplings are switched off as expected. The original
discrete state will generate $2^2$ poles on the four-sheeted Riemann
surface of the analytically continued  $1/\eta(x)$ or
$S$-matrix. When the original discrete state is below the lowest threshold,
it will become a bound state pole on the first sheet and three virtual
state poles on the other three unphysical sheets for small couplings.
When it is above the lowest threshold, it
will move to the other unphysical sheets becoming two pairs of resonance
poles on different sheets. There are also poles on different sheets generated by the
form factors.  From a
mathematical point of view, for each simple pole of the form
factor, there will be $2$ poles of $1/\eta(x)$ generated on the unphysical sheets for
small couplings except for some accidental cases.  All these states can be expressed as the linear
combinations of the original discrete states and continuum states
explicitly. By generalizing the definitions of the continuum states of
\cite{Prigogine:1991},  these discrete states can also enter  the
completeness relation equally. All these discussions can then
be generalized to Friedrichs model with more than two continuum states.
For an $N$-continuum model, the discrete state will generate $2^N$
poles while the number of the poles generated from the form factor
will be $2^{N-1}$. The bare discrete state may be viewed as the bound state
at the more fundamental level such as the bound state of the quarks,
which may be called ``normal states'',
whereas the states generated from the form factors can be viewed as
bound states of the composite level such as the hadrons, which may
represent the so called ``molecular states''. So, we then have a
theoretical criterion
to distinguish the molecular states from the normal states,
that is, the number of poles generated from the molecular state should be one half
of the one for the normal state. This is consistent with the pole counting
rule proposed by Morgan~\cite{Morgan:1992ge}.

The poles on the different sheets with the same origins was noticed  a long time ago by
by Eden and Taylor when studying the property of the $S$-Matrix, and these poles  are
called shadow poles\cite{Eden:1964zz}. However some of the shadow poles are too far away
from the physical region and may not have observable effects in the
experiments. Nevertheless, there could be cases that shadow poles may take
effects in the experiments. In the last two cases in the analysis in
Sect. \ref{sect:polepos},
when $\omega_0$ is near the second threshold, the resonances generated
on the second sheet and third sheet could be close to the physical
region $a_1<\omega<a_2$ and $a_2<\omega$, respectively, and may both have
observable effects in experiments. In fact, some shadow poles may have
already been seen
 in the experiments.  In recent years, we have seen
more and more Quarkonium-like state and exotic state in the
experiments, for example, $X(3872)$, $Z_c(3900)/Z_c(3885)$.
In fact, in \cite{Zhou:2015jta}, a second
sheet pole and a third sheet pole are found to be responsible for the $Z_c(3900)$
and $Z_c(3885)$ line shapes, and both poles originate from a common
bound state. In \cite{zhou:2010ra} the low energy $0^+$ channel, a
dozen of
resonances are found to originate from only four bare states.

We expect that this discussion would be useful in studying the hadron
spectrum, especially on the newly observed exotic states. In fact, the
preliminary attempt has been made recently on the P-wave excited
charmonium states and yielded good results~\cite{Zhou:2017dwj}.
Although our study was motivated by the phenomenology aspects in
hadron interaction, the theory on resonances phenomenon can have much
broader applications.  Actually, the resonance phenomena are observed
in different areas of modern physics, and there are also some variants
of the Friedrichs model in various contexts. The Lee
model~\cite{Lee:1954iq}  is a field theory realization of the
Friedrichs model and there are other second quantized version of the
Friedrichs model like in Refs.~\cite{Antoniou:2003fk,Karpov2000}. In condense matter
physics, the Fano-Anderson~\cite{Fano:1961zz,Anderson:1961} model is
also similar to the Friedrichs model with a bounded continuum
spectrum.  Friedrichs-Fano-Anderson model is also used in atomic
physics, quantum electrodynamics and statistical physics, such as
in the study of the bound state in the continuum phenomena
\cite{Duerinckx1983,PhysRevA.72.063405,LonghiEPJB}. The results  and
the methods in our paper are rather general and  may be helpful in
understanding the diverse phenomena in these areas.

\begin{acknowledgments}
Z.X. is supported by China National Natural
Science Foundation under contract No.  11105138, 11575177 and 11235010.
\end{acknowledgments}

\bibliographystyle{apsrev4-1}%plain, abbrv, alpha

\bibliography{Ref}

\end{document}